\documentclass[aps,showpacs,prb,twocolumn,floatfix]{revtex4-2}
\usepackage{amsmath,amssymb,graphicx,bm}
\newcommand{\Res}{\mathop{\rm Res}\nolimits}
\newcommand{\sgn}{\mathop{\rm sgn}\nolimits}
\newcommand{\ket}[1]{| #1 \rangle}

\newcommand{\avg}[1]{\langle #1 \rangle}

\def\veck{\mathbf k}

\def\be{\begin{equation}}
\def\ee{\end{equation}}

\begin{document}
\title{Many-body perturbation theory for the superconducting quantum dot: Fundamental role of the magnetic field}

\author{V\'aclav  Jani\v{s}}\email{janis@fzu.cz}
\affiliation{Institute of Physics, Academy of Sciences of the Czech
  Republic, Na Slovance 2, CZ-18221 Praha 8, Czech Republic}

\author{ Jiawei Yan} 
\affiliation{Institute of Physics, Academy of Sciences of the Czech Republic, Na Slovance 2, CZ-18221 Praha 8, Czech Republic}

\date{\today}


\begin{abstract}
We develop the general many-body perturbation theory for a superconducting quantum dot represented by a single-impurity Anderson model attached to superconducting leads.  We build our approach on a thermodynamically consistent mean-field approximation with a two-particle self-consistency of the parquet type. The two-particle self-consistency leading to a screening of the bare interaction proves substantial for suppressing the spurious transitions of the  Hartree-Fock solution.   We demonstrate that the magnetic field plays the fundamental role in the extension of the perturbation theory beyond the weakly correlated $0$-phase. It controls the critical behavior of the $0-\pi$ quantum transition, lifts the degeneracy in the $\pi$-phase, where the limits to zero temperature and zero magnetic field do not commute. The response to the magnetic field is quite different  in $0$- and $\pi$-phases. While the magnetic susceptibility vanishes in the $0$-phase it becomes of the Curie type and diverges in the $\pi$-phase at zero temperature.      
\end{abstract}

\maketitle 

\section{Introduction}
\label{sec:intro}

Nanostructures with well separated localized energy levels are objects that can be isolated in regions of a few nanometers or microns.  They can be experimentally realized by magnetic impurities on metallic surfaces \cite{Ralph:1994aa,Madhavan:1998aa,Li:1998aa,Goldhaber:1998aa,Goldhaber-Gordon:1998aa,Cronenwett:1998aa},  semiconducting  quantum dots \cite{Katsaros10} nanowires \cite{vanDam06,Jesper13},  carbon nanotubes \cite{Kasumov99,Morpurgo99,Kasumov03,Jarillo06,Jorgensen06,Cleuziou06,Jorgensen07,Grove07,Pallecchi08,Zhang08,Jorgensen09,Eichler09,Liu09,Pillet10,Lee12,Maurand12,Pillet13,Kumar14,Delagrange15} or single $C_{60}$ molecules \cite{Winkelmann09}.  They are ideal systems for studying elementary quantum mechanical phenomena according to the substrates on which they are grown or in which they are embedded due to a detailed control of the relevant microscopic parameters. When the impurity atoms with unpaired correlated electrons are placed in metals one observes the Kondo effect \cite{Goldhaber:1998aa,Goldhaber-Gordon:1998aa,Cronenwett:1998aa}. The correlated quantum nanostructres attached to superconductors represent  tunable microscopic Josephson junctions \cite{Kasumov99,Kasumov03,DeFranceschi10}. The simultaneous presence of strong electron correlations on semiconducting impurities and proximity of superconductors allow us to observe and analyze the interplay between the Kondo effect and the formation of the Cooper pairs carrying the Josephson current through the semiconducting nanodevices \cite{Matsuura77,Glazman89,Rozhkov:2000aa,Buitelaar02,Aono:2004aa,Graber04,Siano04,Choi04,vanDam06,Cleuziou06,Jorgensen07,Tanaka07,Grove07,Lim08,Karrasch08,Eichler09,Yamada:2010aa,Yamada:2011aa,Luitz12,Oguri13}.  

Strong Coulomb repulsion on quantum dots attached to superconducting leads may cause a local quantum critical point at which  the lowest many-body eigenstates of the system cross and  a spin-singlet ground state with the positive supercurrent ($0$-phase) goes over to a spin-doublet state with a small negative supercurrent ($\pi$-phase) \cite{Matsuura77,Glazman89,Rozhkov:2000aa,Yoshioka00,Siano04,Choi04,Sellier05,Novotny05,vanDam06,Cleuziou06,Jorgensen07,Karrasch08,Meng:2009aa,Luitz12,Maurand12}.  This transition is associated with crossing of the Andreev bound states (ABS) at the Fermi energy as has also been observed experimentally \cite{Pillet10,Pillet13,Jesper13}. 
 
A number of theoretical techniques have been used to address the $0-\pi$ transition and related properties of superconducting quantum dots. A very good quantitative agreement with the experiments \cite{Luitz12,Pillet13,Delagrange15} can be obtained in a wide range of parameters using heavy numerics such as numerical renormalization group (NRG) \cite{Yoshioka00,Choi04,Bauer07,Tanaka07,Hecht08,Oguri04,Rodero12,Oguri13,Pillet13} and quantum Monte Carlo (QMC) \cite{Siano04,Luitz10,Luitz12,Delagrange15,Pokorny:2018aa}. However, both NRG and QMC demand extensive time and computational resources and they do not disclose the microscopic origin of this quantum critical behavior.  They are also unable to distinguish the physically different properties of the in-gap states in the $0$- and $\pi$-phases. Alternatively, analytic approaches have been used mostly based on perturbation expansions, either in the strength of the Coulomb repulsion  \cite{Alastalo:1998aa,Vecino:2003aa,Zonda:2015aa,Janis:2016aa,Zonda:2016aa,Domanski:2017aa}  or around the atomic limit  \cite{Konig08,Meng09,Droste12,Wentzell:2016aa}.
 
The perturbation expansion in a small parameter cannot describe any collective behavior. A self-consistent summation of infinite series must be included to interpolate between weak and strong couplings needed to describe the $0-\pi$ transition.  Summations via self-consistences are used both in the expansion in the Coulomb repulsion and around the atomic limit \cite{Meden:2019aa}.    Perturbation expansions around the atomic limit miss the strong-coupling Kondo effect for narrow superconducting gaps. The expansion in the Coulomb repulsion is well defined only at zero temperature and in the weakly-coupled spin-symmetric state of the $0$-phase. The standard way to include critical behavior and phase transitions is to use a mean-field approximation with spin-polarized states as a starting point for the perturbation expansion \cite{Vecino:2003aa}. Although the mean-field, Hartree-Fock approximation may give reasonably good quantitative predictions for weak and moderate coupling \cite{Martin:2012aa} it is conceptually unacceptable, since the real $0-\pi$ transition is a consequence of a spurious critical transition to the magnetic state \cite{Meden:2019aa}.  

There is a way to improve upon the improper start of the perturbation expansion in the interaction strength. One has to replace the weak-coupling mean-field approximation with an advanced one that is able to interpolate consistently between the weak and strong couplings. It must be a self-consistent theory suppressing the spurious transition to the magnetic state and reproducing the Kondo strong-coupling regime in the impurity models. One of the authors proposed a mean-field theory with a two-particle self-consistency that is free of any unphysical behavior and qualitatively correctly reproduces the Kondo limit of the single-impurity Anderson model (SIAM)  \cite{Janis:2007aa,Janis:2008ab,Janis:2017aa,Janis:2017ab,Janis:2019aa}.  This mean-field approximation is a thermodynamically consistent extension of the weak-coupling theory to the whole range of the input parameters. 
 
It is the aim of this paper to apply the mean-field approximation from references \cite{Janis:2007aa,Janis:2008ab,Janis:2017aa,Janis:2017ab,Janis:2019aa} on the Anderson impurity attached to superconducting leads. The superconducting leads induce a gap on the impurity with no states at the Fermi energy. Instead, discrete in-gap states emerge the position of which depends on the interaction strength and the phase difference between the attached superconducting electrodes.  The theory developed for the SIAM with non-zero density of states at the Fermi energy will  be appropriately modified to offer a reliable description of  the models with a gap.  The extension of the mean-field approach to the singlet phase of the superconducting quantum dot seems straightforward, since it is the ground state in weak coupling. An extension to the doublet phase with a degenerate ground state and no weak-coupling regime  appears to be more elaborate.  
  
The many-body perturbation theory  for low-energy excitations can be used only with a unique, non-degenerate many-body ground state.  It means that a degeneracy of the ground state in the doublet phase must be lifted before we can apply the many-body Green function technique and the diagrammatic expansion. The doublet ground state is degenerate with respect to the spin reflection. We must then use a small magnetic field on the impurity to lift the degeneracy. We hence need to formulate the mean-field approximation for the dot in an external magnetic field. The properties of the superconducting quantum dot with a Zeeman field were recently  studied experimentally \cite{Cornils:2017aa,Dvir:2019aa,Corral:2020aa,Whiticar:2021aa} and also theoretically \cite{Tanaka:1995aa,Yamada:2007aa,Li:2014aa,Kirsanskas:2015aa,Zitko:2015aa,Wentzell:2016aa,Gerven-Oei:2017aa,Lim:2020aa}.  The role of the Zeeman field in the perturbation theory of the superconducting quantum dot is crucial.  It allows us to circumvent the quantum critical point of the $0-\pi$ transition and to extend the many-body approach from weak to strong coupling regimes at zero temperature. The limits to zero filed and zero temperature do not commute in the $\pi$-phase. Moreover, the response to the magnetic field is crucial for distinguishing between the $0$-phase with bound singlet Cooper pairs and the $\pi$-phase with in-gap fermionic excitations carrying a local magnetic moment. This will be demonstrated on the behavior of the magnetic susceptibility.  This feature has not yet been disclosed because a full consistent many-body theory of the superconducting quantum dot with the Zeeman field and at arbitrary temperature is still missing.  
  
The layout of the paper is the following. We introduce the model and the Nambu formalism of the correlated impurity attached to superconducting leads in Sec. II. We introduce the basic ingredients of standard many-body perturbation theory in Sec. III. The core of our thermodynamically consistent mean-field approximation with a two-particle self-consistency is presented in Sec. IV.  We apply our mean-field approximation to study the behavior of the in-gap states and the $0-\pi$ transition in Sec. V. Explicit calculations are performed in the asymptotic atomic limit of the infinite superconducting gap in Sec. VI. Numerical results are presented in Sec. VII and Sec. VIII brings concluding remarks. Less important and elucidating technical details are presented in Appendices A-C.

\section{Model Hamiltonian and Andreev bound states}
\label{sec:model}

Standardly, a single impurity is used to simulate the nanowire with separated energy levels connecting superconducting leads in the experimental setup. The Hamiltonian of the system consisting of a single impurity attached to BCS superconducting leads is
\begin{equation}
\mathcal{H}=\mathcal{H}_{dot}+\sum_{s=R,L}(\mathcal{H}^s_{lead}+\mathcal{H}^s_c)\,,
\end{equation}
where the impurity Hamiltonian is a single-level atom with the level energy $\pm\epsilon$ for single electron (hole) and Coulomb repulsion $U$ in the Zeeman magnetic field $h$
\begin{equation}
\mathcal{H}_{dot}=\sum_{\sigma=\pm 1}\left(\epsilon - \sigma h\right)  d_\sigma^\dagger d_\sigma^{\phantom{\dag}}
+Ud_\uparrow^\dag d_\uparrow^{\phantom{\dag}} d_\downarrow^\dag d_\downarrow^{\phantom{\dag}} \,,
\end{equation}
where $\sigma =\pm 1$ corresponds to spin up/down (parallel/antiparallel to the applied magnetic field).
 
The Hamiltonian of the leads is 
\begin{multline}
\mathcal{H}^s_{lead}=\sum_{\mathbf{k}\sigma}
\epsilon(\mathbf{k})c_{s\mathbf{k}\sigma}^\dag c_{s\mathbf{k}\sigma}^{\phantom{\dag}} \\
-\Delta_{s}\sum_\mathbf{k}(e^{i\Phi_s}
c_{s\mathbf{k}\uparrow}^\dag c_{s\mathbf{-k}\downarrow}^\dag+\textrm{H.c.})\end{multline}
with $s = L,R$ denoting left, right lead. Finally, the hybridization term for the contacts reads
\begin{equation}
\mathcal{H}^s_c=-t_s\sum_{\mathbf{k}\sigma}
(c_{s\mathbf{k}\sigma}^\dag d_\sigma^{\phantom{\dag}}+\textrm{H.c.}) \ .
\end{equation}

We use the Nambu spinor formalism to describe the Cooper pairs and the anomalous functions related with the superconducting order parameters and breaking charge conservation.  The Nambu spinors in the superconducting leads are 
\begin{equation}
\widehat{\varphi}^{\phantom{\dagger}}_{s\mathbf{k}\sigma} = 
\begin{pmatrix}
c^{\phantom{\dagger}}_{s\mathbf{k}\sigma} \\ c^{\dagger}_{s\bar{\mathbf{k}}\bar{\sigma}}
\end{pmatrix} \quad , \quad  \widehat{\varphi}^{\dagger}_{s\mathbf{k}\sigma} = \begin{pmatrix}c^{\dagger}_{s\mathbf{k}\sigma} & c^{\phantom{\dagger}}_{s\bar{\mathbf{k}}\bar{\sigma}}\end{pmatrix}\,,
\end{equation}
where we introduced $\bar{\veck}= - \veck$ and $\bar{\sigma} = -\sigma$.

Due to the hybridization the Cooper pairs can penetrate onto the impurity giving rise to anomalous impurity Green functions. Hence, we introduce the Nambu spinors also for the impurity (local) operators 
\begin{equation}
\widehat{\phi}^{\phantom{\dagger}}_{\sigma} = 
\begin{pmatrix}
d^{\phantom{\dagger}}_{\sigma} \\ d^{\dagger}_{\bar{\sigma}}
\end{pmatrix} \quad , \quad  \widehat{\phi}^{\dagger}_{\sigma} = \begin{pmatrix}d^{\dagger}_{\sigma} & d^{\phantom{\dagger}}_{\bar{\sigma}}\end{pmatrix}\,.
\end{equation}

The individual degrees of freedom of the leads are unimportant for the impurity quantities and we integrate them out leaving only the impurity variables dynamical. The fundamental function  after projecting the lead degrees of freedom is the one-electron impurity Green function measuring (imaginary) time fluctuations that in the Nambu formalism is a $2\times 2$ matrix  
\begin{multline}
\widehat{G}_{\sigma}(\tau-\tau') \\=-
\begin{pmatrix}
\langle \mathbb T\left[d_\sigma(\tau) d_\sigma^{\dag}(\tau')\right]\rangle\ , & \langle \mathbb T\left[ d_{\sigma}(\tau) d_{\bar{\sigma}}(\tau')\right]\rangle \\[0.3em]
\langle \mathbb T\left[d_{\bar{\sigma}}^{\dag}(\tau) d_{\sigma}^{\dag}(\tau')\right]\rangle \ , & \langle \mathbb T\left[d_{\bar{\sigma}}^\dag(\tau) d_{\bar{\sigma}}(\tau')\right]\rangle
\end{pmatrix} \\ 
=
\begin{pmatrix}
G_{\sigma}(\tau - \tau')\ ,		& 	\mathcal{G}_{\sigma}(\tau - \tau') 	\\
\bar{\mathcal{G}}_{\sigma}(\tau - \tau') \ ,	& 	\bar{G}_{\sigma}(\tau - \tau')
\end{pmatrix} 
\end{multline}
correlating appearance of electrons and holes with specific spin on the impurity.  
We introduced normal particle  and hole $G_{\sigma}, \bar{G}_{\sigma}$ propagators that conserve spin and anomalous  $\mathcal{G}_{\sigma}$  and $\bar{\mathcal{G}}_{\sigma}$ Green functions that create and annihilate singlet Copper pairs in the spin-polarized solution. 

The electron and hole functions  are connected by symmetry relations $\bar{G}_{\sigma}(\tau) = - G_{\bar{\sigma}}(-\tau) = - G_{\bar{\sigma}}(\tau)^{*}$,  $\bar{\mathcal{G}}_{\sigma}(\tau) = \mathcal{G}_{\bar{\sigma}}(-\tau)^{*}$, and $\mathcal{G}_{\sigma}(\tau) = -\mathcal{G}_{\bar{\sigma}}(-\tau)$.   

The problem can be exactly solved for an impurity without the onsite interaction, $U=0$. In this case the inverse unperturbed propagator for the spin-polarized situation can be represented in the Nambu formalism as a matrix.  We use identical left and right hybridizations to superconductors, $t_{L} = t_{R} = t$ without loss of generality. The asymmetric situation can be transformed to a symmetric one \cite{Kadlecova:2017aa}.  Due to energy conservation it is convenient to use Fourier transform from (imaginary) time to frequency (energy) where the Green function can analytically be continued to complex values.  The matrix of the inverse Green function for a complex energy $z$ reads
\begin{equation}
\widehat{G}_{\sigma}^{-1}(z)= 
\begin{pmatrix}
z[1+s(z)] + \sigma h  - \epsilon\,, & \Delta\cos(\Phi/2)s(z) \\[0.3em]
\Delta\cos(\Phi/2)s(z)\,, & z[1+\sigma(z)] + \sigma h + \epsilon
\end{pmatrix},
\end{equation}
where 
\begin{equation}
s(z)=\frac{i\Gamma_0}{\zeta}\mathrm{sgn}(\Im z).
\end{equation}
is the ``hybridization self-energy'' $\sigma(z)$, that is, a dynamical renormalization of the impurity energy level due to the hybridization to the superconducting leads. We approximated the Green function in the leads by its value  at the Fermi energy and denote $\Gamma_{0}= 2\pi t^{2}\rho_{0}$ being the effective hybridization strength. We denoted $\Phi = \Phi_{L} - \Phi_{R}$ the difference between the phases of the attached superconducting leads and $\rho_{0}$ the density of states of the lead electrons at the Fermi energy.   To represent explicitly the hybridization self-energy we introduced a new complex number $\zeta=\xi+i\eta$ derived from the complex energy $z = x + iy$ by a quadratic equation $\zeta^2=z^2-\Delta^2$. Thereby the following convention for the complex square root has been used 
\begin{equation}
\xi\eta=xy, \quad \mathrm{sgn}(\xi) = \mathrm{sgn} (x), \quad \mathrm{sgn}(\eta) = \mathrm{sign} (y)\ .
\end{equation}

The renormalized energy $\zeta$  along the real axis $z=x\pm i0$ is real outside the energy gap $(-\Delta,\Delta)$ and imaginary within it
\begin{equation}
\begin{split}
\zeta&=\sgn(x)\sqrt{x^2-\Delta^2}\qquad\mathrm{ for }\qquad |x|>\Delta,\\
\zeta&=\pm i\sqrt{\Delta^2-x^2}\qquad\qquad \!\! \mathrm{ for }\qquad |x|<\Delta \ .
\end{split}
\end{equation}
Accordingly the hybridization self-energy is purely imaginary outside the gap and real within it
\begin{equation}
\begin{split}
s(x\pm i0) &=\pm\frac{i\Gamma_0\sgn(x)}{\sqrt{x^2-\Delta^2}}\qquad\mathrm{ for }\qquad|x|>\Delta\ ,
\\
s(x\pm i0) &=\phantom{\pm}\frac{\Gamma_0}{\sqrt{\Delta^2-x^2}}\qquad\mathrm{ for }\qquad|x|<\Delta\ .
\end{split}
\end{equation}

With the above definitions the unperturbed ($U=0$) impurity Green function is
\begin{multline}\label{eq:D0}
\widehat{G}^{(0)}_{\sigma}(z)=\frac{1}{D_{\sigma}(z)}
\\
\times
\begin{pmatrix}
z[1+s(z)] + \sigma h + \epsilon\ , & - c_{\Phi}\Delta s(z) \\[0.3em]
- c_{\Phi}\Delta s(z)\ , & z[1+s(z)] + \sigma h - \epsilon
\end{pmatrix}.
\end{multline}
where we denoted $c_{\Phi}=\cos(\Phi/2) $ and introduced
$$
D_{\sigma}(z) = \left[z(1+s(z)) + \sigma h\right]^{2} - \epsilon^{2} - c_{\Phi}^{2}\Delta^{2}s(z)^{2} 
$$
the determinant of the matrix of the inverse unperturbed impurity Green function. It is decisive for the determination of the gap states. This determinant is real within the gap and can goes through zero determining the gap states that are simultaneously the Andreev states. They are four of them $\pm \omega_{\sigma}$ in the external magnetic field.  We denote the two independent 
\begin{equation}
\omega_{\sigma}(1+ s_{\sigma}) = - \sigma h \pm \sqrt{\epsilon^{2} + c_{\Phi}^{2}\Delta^{2}s_{\sigma}^{2}} \,.
\end{equation}
where we used Eq.~\eqref{eq:D0} and denoted $s_{\sigma}=s(\omega_{\sigma})$.

\section{Perturbation expansion: Diagrammatic representation}
\label{sec:PE}

The best way to represent the many-body perturbation expansion is to use a graphical, diagrammatic representation that can be introduced also in the Nambu formalism. We start with the diagrammatic representation of the Nambu spinor of the impurity propagator to which we assign solid lines decorated with arrows as follows 
\begin{multline}
\begin{pmatrix}
G_{\sigma}(i\omega_{n})\ ,		&  \mathcal{G}_{\sigma}(i\omega_{n}) 	\\
\bar{\mathcal{G}}_{\sigma}(i\omega_{n}) \ ,	& 	\bar{G}_{\sigma}(i\omega_{n}) 
\end{pmatrix}
\\=
\begin{pmatrix}
G_{\sigma}(i\omega_{n})\ ,		&  \mathcal{G}_{\sigma}(i\omega_{n}) 	\\
\mathcal{G}_{\bar{\sigma}}^{*}( -i\omega_{n}) \ ,	&  - G_{\bar{\sigma}}( - i\omega_{n}) 
\end{pmatrix}
\\ =
\begin{pmatrix}
  \quad\includegraphics{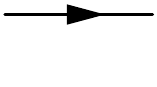}\quad & \quad\includegraphics{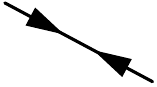}\quad \\
 \quad\includegraphics{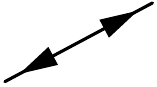}\quad & \quad\includegraphics{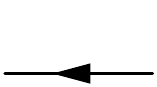}\quad\end{pmatrix}
\end{multline}
We used the symmetry relations of the unperturbed Green functions that remain generally valid and read in (complex) energy representation
\begin{subequations}
\begin{align}\label{eq:EH-G}
\bar{G}_{\sigma}(i\omega_{n}) &= - G_{\bar{\sigma}}(-i\omega_{n}) = - G_{\bar{\sigma}}^{*}(i\omega_{n})\,, \\
\bar{\mathcal{G}}_{\sigma}(i\omega_{n})& =\mathcal{G}_{\bar{\sigma}}^{*}(-i\omega_{n}) =  - \mathcal{G}_{\bar{\sigma}}^{*}(i\omega_{n})\, .
\end{align}
\end{subequations}

We keep the time (charge) propagation (from left to right)  in the diagrammatic representation and attach the spin up/down to the upper/lower line. Anomalous propagators do not conserve charge by annihilating two electrons with opposite spins (arrows against each other) or create a Cooper pair (arrows from each other). We can construct standard Feynman many-body diagrams for processes induced by the Coulomb interaction of the electrons on the impurity between two superconducting leads. The Coulomb interaction will be represented via a wavy line. Since the interaction is static, the interaction wavy line is always vertical. Before we start to analyze the diagrammatic contributions from the perturbation expansion we resume basic exact relations.  

The impact of the Coulomb repulsion on the one-electron Green function is included in a matrix self-energy $\hat{\Sigma}(z)$ so that the full inverse propagator in the spin-polarized situation reads 
$\widehat{G}^{-1}(i\omega_{n}) =\widehat{G}_0^{-1}(i\omega_{n})-\widehat{\Sigma}(i\omega_{n})$. Its explicit component representation is
%
\begin{multline}
\widehat{G}_{\sigma}(i\omega_{n})
\\
=\frac{1}{D_{\sigma}(i\omega_{n})}
\begin{pmatrix}
- X_{\bar{\sigma}}(-i\omega_{n}) , &  -c_{\Phi}Y(i\omega_{n}) \\[0.3em]
-c_{\Phi} Y^{*}(-i\omega_{n}) , & X_{\sigma}(i\omega_{n})
\end{pmatrix}
\end{multline}
 with
\begin{subequations}\label{eq:XY-full}
\begin{multline}
X_{\sigma}(i\omega_{n}) = i\omega_{n}[1+s(i\omega_{n})]  + \sigma\left[ h - \Delta\Sigma(i\omega_{n})\right]
\\
 - \epsilon - \Sigma(i\omega_{n}) \,,
\end{multline}
\begin{align}
Y(i\omega_{n}) &= s(i\omega_{n})\Delta   - \mathcal{S}(i\omega_{n})\,.
\end{align}
\end{subequations}
We denoted $\Sigma(i\omega_{n})$, $\Delta\Sigma(i\omega_{n})$  the even and odd parts of the normal self-energy with respect to the magnetic field  and $\mathcal{S}(i\omega_{n})$ the anomalous superconducting part of the interaction-induced self-energy. The even, spin-symmetric self energy, $\Sigma(i\omega_{n})$ and the anomalous one, $\mathcal{S}(i\omega_{n})$, will be determined form the dynamical spin-symmetric Schwinger-Dyson equation. The odd self-energy $\Delta\Sigma(i\omega_{n})$ generalizes the classical order parameters and will be related with the two-particle irreducible vertex via a linearized Ward identity \cite{Janis:2019aa}.  
 
The spin-dependent determinant of the inverse of the matrix propagator in this notation is 
 \begin{multline}\label{eq:D-Sigma}
D_{\sigma }(i\omega_{n})
\\
= - X_{\sigma}(i\omega_{n}) X_{\bar{\sigma}}(-i\omega_{n}) - c_{\Phi}^{2} Y(i\omega_{n})Y^{*}(-i\omega_{n}) \,,
\end{multline}
with the electron-hole symmetry $D_{\sigma }(i\omega_{n}) = D_{-\sigma }(-i\omega_{n})$.

The normal spin-dependent impurity propagators are
\begin{subequations}
\begin{align}
G_{\sigma}(i\omega_{n}) &=  - \frac{X_{\bar{\sigma}}(-i\omega_{n})}{D_{\sigma}(i\omega_{n})} \,,\\ 
\bar{G}_{\sigma}(i\omega_{n}) &=   \frac{X_{\sigma}(i\omega_{n})}{D_{\sigma}(i\omega_{n})} \,,
\end{align}
\end{subequations}
while the anomalous propagators are
\begin{subequations}
\begin{align}
\mathcal{G}_{\sigma}(i\omega_{n}) &=  - c_{\Phi}\frac{Y(i\omega_{n})}{D_{\sigma}(i\omega_{n})}\,,\\  
\bar{\mathcal{G}}_{\sigma}(i\omega_{n}) &=  - c_{\Phi}\frac{Y^{*}(-i\omega_{n})}{D_{\sigma}(i\omega_{n})}\,.
\end{align}
\end{subequations}

The existence and positions of the Andreev states are again determined from zeros of determinant $D_{\sigma}(i\omega_{n})$. They depend on the behavior of the normal and anomalous self-energies for which we introduce a diagrammatic expansion.  We first formulate the perturbation expansion in the thermodynamic language using the Matsubara representation. Only after having constructed contributions to the perturbation expansion and within the selected approximations we perform analytic continuation to real frequencies so that to control the behavior of the Andreev bound states (ABS).

\section{Perturbation expansion: Reduced parquet equations}
 
The basic element of the many-body perturbation expansion is the one-particle propagator. Knowing it we determine all the physical quantities. The Dyson equation introduces the self-energy containing the whole impact of the particle interactions on the one-particle propagator. That is why most of the theoretical approaches focus on the self-energy.  It is, however, not the best way to control the critical  and crossover behavior from weak to strong coupling regimes. Although it is more elaborate and complex in its analytic structure, perturbation theory applied directly to two-particle functions has gained on popularity in recent years. The idea to extend the perturbation theory and its renormalizations to two-particle functions is rather old \cite{DeDominicis:1964aa,DeDominicis:1964ab}. Presently, this general approach is used within the so-called parquet equations that add a two-particle self-consistency \cite{Bickers:1991aa,Rohringer:2018aa}. Generally, the full unrestricted approximations at the two-particle level can be solved only numerically and in the Matsuubara formalism at non-zero temperatures. One has to resort to simplifications if the  critical behavior should be controlled analytically. We developed the so-called reduced parquet equations to reach this objective \cite{Janis:2007aa,Janis:2008ab,Janis:2017aa,Janis:2017ab,Janis:2019aa}. The fundamental idea of this two-particle approach is to treat approximate two-particle vertex functions and one-particle self-energy separately and match them at the end so that to keep the theory thermodynamically consistent and conserving.           
 
\subsection{Two-particle vertex: Effective interaction}
%
\begin{figure*}
\includegraphics[width=15cm]{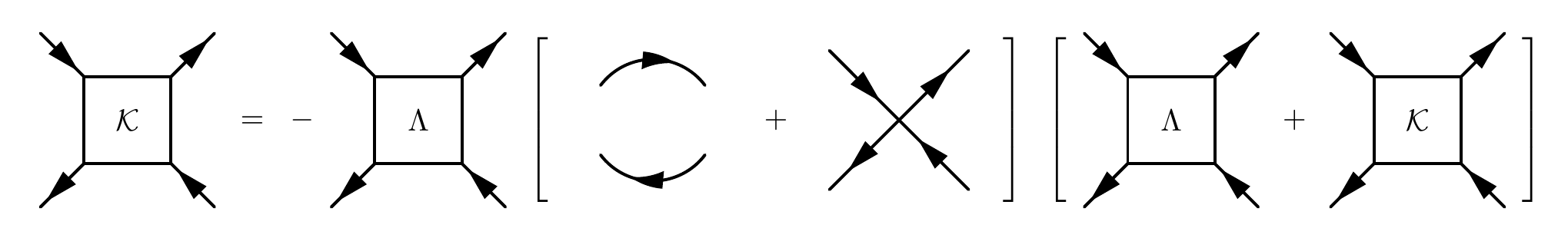}
\caption{Diagrammatic representation of the Bethe-Salpeter equation for the reducible vertex in the electron-hole channel. The electron-hole propagator contains simultaneous normal and anomalous propagators. The lines from the central part in the brackets are attached to the left and right vertices (the three parts separated by brackets are mathematically multiplied) to form two connected diagrams. \label{fig:RPE-EH-SC}}
\end{figure*} 
%
\begin{figure}
\includegraphics[width=8cm]{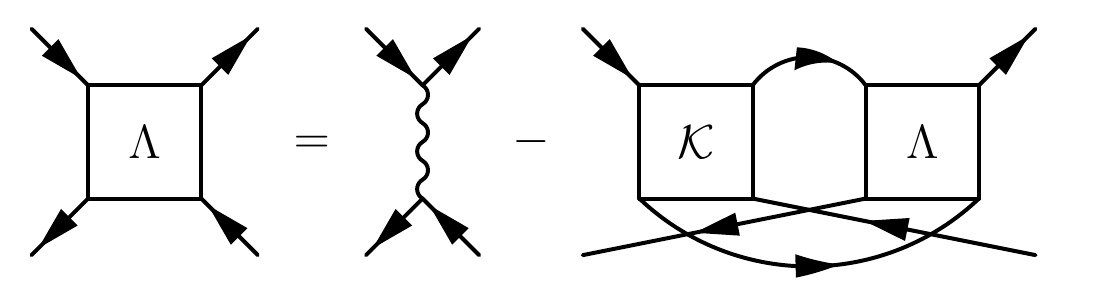} 
\caption{The reduced Bethe-Salpeter equation as explained in the text  for the irreducible vertex from the electron-hole scattering channel. The electron-electron propagator does not contain an anomalous part due to conservation laws. \label{fig:RPE-EE-SC}}
\end{figure} 

The fundamental element in the two-particle perturbation theory is the two-particle vertex $\Gamma$. It has generally three dynamical variables, two fermionic $i\omega_{n},i\omega_{n'}$, one bosonic $i\nu_{m}$, and two spin indices $\sigma,\sigma'$. An irreducible vertex $\Lambda$ plays the role of the two-particle self-energy. The two-particle irreducibility is not uniquely defined and hence there is not a unique way to select the irreducible vertex \cite{Bickers:1991aa}. The most important one is, however, that from the two-particle scattering channel leading to a singularity and a critical behavior in intermediate coupling. It is the spin-singlet  electron-hole scattering channel. The full vertex then can be decomposed into its irreducible $\Lambda$ and reducible, $\mathcal{K}$, parts  
\begin{multline}\label{eq:Gamma-full}
\Gamma_{\uparrow\downarrow}(i\omega_{n},i\omega_{n'};i\nu_{m}) = \Lambda_{\uparrow\downarrow}(i\omega_{n},i\omega_{n'};i\nu_{m})  
\\
\ + \mathcal{K}_{\uparrow\downarrow}(i\omega_{n},i\omega_{n'};i\nu_{m}) \,,
\end{multline}
where $\omega_{n}$ and $\omega_{n'}$ are energies of the incoming and outgoing electron, respectively, and $\nu_{m}$ is the energy difference between the electron and the hole that is conserved in the multiple singlet electron-hole scatterings. 

Generally, the reducible vertex in one scattering channel becomes irreducible in the other scattering channels. The parquet equations self-consistently intertwine them to determine both irreducible and reducible parts of the full vertex. Our approximation resorts to a two-channel version of the parquet equations with only singlet electron-hole and electron-electron multiple scatterings. The sum of the series of the repeated scatterings of particle pairs are mathematically  represented by the Bethe-Salpeter equations. The Bethe-Salpeter equation in the electron-hole channel  determines the reducible vertex $\mathcal{K}$ as a functional of the irreducible one $\Lambda$ and is diagrammatically represented in Fig.~\ref{fig:RPE-EH-SC}. The irreducible vertex in our approximation is determined from a reduced Bethe-Salpeter equation that is diagrammatically represented in Fig.~\ref{fig:RPE-EE-SC}. The reduction of the full Bethe-Salpeter equation in the electron-electron channel consists in suppressing convolution of two diverging reducible vertices $\mathcal{K}GG\mathcal{K}$ so that not to destroy the possible quantum criticality in the strong-coupling regime of the full vertex $\Gamma$. The suppressed term is expected to be  compensated by higher-order terms not included in the two-channel approximation \cite{Janis:2019aa}. 

 The mean-field approximation enters these reduced parquet equations by replacing the irreducible vertex by a frequency and spin-independent constant $\Lambda$ that then plays a role of an effective interaction. The reducible vertex determined by the equation of Fig~\ref{fig:RPE-EH-SC}  is 
\begin{equation}
\mathcal{K}_{\sigma}(i\nu_{m})= -\frac{\Lambda^{2}\phi_{\sigma}(i\nu_{m})}{1 + \Lambda\phi_{\sigma}(i\nu_{m})} \,,
\end{equation} 
 where the fermionic frequencies are $\omega_{n}=(2n + 1)\pi T$ and $\omega_{n'}=(2n' + 1)\pi T$, and the bosonic is  $\nu_{m}= 2m\pi T$. We denoted the full electron-hole bubble
\begin{multline}
\phi_{\sigma}(i\nu_{m}) = \frac 1\beta\sum_{\omega_{n}}\left[G_{\bar{\sigma}}(i\omega_{n} + i\nu_{m}) G_{\sigma}(i\omega_{n}) 
\right. \\ \left.
+\ \mathcal{G}_{\bar{\sigma}}(i\omega_{n} + i\nu_{m}) \mathcal{G}_{\sigma}(i\omega_{n}) \right] \,.
\end{multline}  
 
 The reduced parquet equations are justified in the critical region of the magnetic transition, that is, in the spin symmetric case where $G_{\uparrow} = G_{\downarrow}$. The mean-field approximation must be, however, defined in the whole representation space, including the spin-polarized state. Since we introduced only a spin-independent renormalization of the bare interaction strength, we replace the spin-dependent bubble with its symmetric form, that is, $\phi_{\sigma}(i\nu_{m}) \to  \phi(i\nu_{m}) = (\phi_{\uparrow}(i\nu_{m}) + \phi_{\downarrow}(i\nu_{m}))/2$ to determine the effective interaction $\Lambda$. Inserting this function into the reduced Bethe-Salpeter equation from Fig~\ref{fig:RPE-EE-SC} leads to
 \begin{multline}\label{eq:Lambda-full}
\left[1 + \frac 1\beta\sum_{\nu_{m}} \mathcal{K}(-i\nu_{m})G_{\uparrow}(i\omega_{n + m})
\right. \\ \left. \phantom{\frac 12}
\times  G_{\downarrow}(i\omega_{n'- m}) \right] \Lambda = U \,,
\end{multline}
which cannot, however, be satisfied for all fermionic frequencies. An approximate treatment of this equation is necessary to close the mean-field scheme. 

The dominant contribution in metallic systems to vertex $\Lambda$ comes from the lowest Matsubara frequencies close to the Fermi energy, that is $|n|\approx|n'|\approx 0$.  We can then take the lowest values near the Fermi energy at low-temperatures as we did in the SIAM \cite{Janis:2017aa,Janis:2017ab,Janis:2019aa}. The Fermi energy of the superconducting quantum dot lies in the gap and there is no contribution from small fermionic frequencies to screening of the interaction.  The fluctuations in the fermionic Matsubara frequencies may shift the value of the critical interaction but do not affect the universal critical behavior. We can use averaging over the fermionic Matsubara frequencies to obtain a mean-field (static) renormalization of the bare interaction strength at any temperature within the same universality class \cite{Janis:2007aa}.  The averaging is not uniquely defined and the optimal one, producing the most accurate result, depends on the studied problem. We found that the  most suitable averaging scheme here is to multiply Eq.~\eqref{eq:Lambda-full} by a product $G_{\uparrow}(-i\omega_{n'}) \exp(-i\omega_{n'}0^{+})G_{\downarrow}(-i\omega_{n})\exp(-i\omega_{n}0^{+})$ and sum over the fermionic frequencies. The resulting equation for the effective interaction $\Lambda$ then is
\begin{equation}\label{eq:Lambda-general}
\Lambda =   \frac {U n_{\uparrow}n_{\downarrow}  }{ n_{\uparrow}n_{\downarrow}  + \Lambda^{2} \mathcal{X}} 
\end{equation}
where $n_{\sigma}$ is the density electrons with spin $\sigma$ and 
\begin{equation}\label{eq:X-integral}
\mathcal{X} =  -\frac {1}{\beta} \sum_{\nu_{m}}\frac{ \psi(i\nu_{m}) \psi(-i\nu_{m})\phi(-i\nu_{m})}{1 + \Lambda\phi(-i\nu_{m})} \,.
\end{equation}
 We introduced the electron-electron bubble 
\begin{align}
\psi(i\nu_{m}) &=  \frac 1{\beta } \sum_{\omega_{n}} G_{\downarrow}(i\omega_{m + n})G_{\uparrow}(-i\omega_{n}) 
\nonumber\\
&=  \frac 1{\beta} \sum_{\omega_{n}} G_{\uparrow}(i\omega_{m + n})G_{\downarrow}(-i\omega_{n}) \,,
\end{align}
 which is spin independent.
 
Equation~\eqref{eq:Lambda-general} determines the effective interaction for the known densities $n_{\sigma}$ and the screening integral $\mathcal{X}$. The explicit solution for $\Lambda$ can be obtained by a substitution with an auxiliary variable $w$  
\begin{subequations}\label{eq:Lambda-cubic}
\begin{equation}\label{eq:Lambda-w}
\Lambda = w - \frac{n^{2} - m^{2}}{12w\mathcal{X}} \,,
\end{equation} 
where we used $n_{\sigma} = (n + \sigma m)/2$ with $n$ and $m$ being the total charge and spin density, respectively. The cube of the new variable $w^{3}$ satisfies a quadratic equation with a single positive root
 \begin{equation}\label{eq:w3-roots}
 w^{3} = \frac{U\left(n^{2} - m^{2}\right)}{8\mathcal{X}}\left[1 + \sqrt{1 + \frac 1{27}\ \frac{n^{2} - m^{2}}{U^{2}\mathcal{X}}} \right]\,.
\end{equation}  
\end{subequations}
The known value of $w$ determined the effective interaction $\Lambda$ from Eq.~\eqref{eq:Lambda-w}. The consistency condition for positivity of the effective interaction is
\be
12w^{2}\mathcal{X} \ge n^{2} - m^{2}\,.
\ee
Equation~\eqref{eq:Lambda-cubic} does not, however, determine the effective interaction  explicitly since integral $\mathcal{X}$ depends on the solution. The final solution can be reached only via iterations.

\subsection{Thermodynamic propagators: Static self-energies} 

One cannot close the equation for the two-particle vertex $\Lambda$ without connecting it with the one-particle densities. It means that we must determine how the self-energy in the one-particle propagators determining the charge and spin densities is related with the two-particle vertex from Eq.~\eqref{eq:Lambda-general}.  We introduce two self-energies according to their symmetry with respect to the spin reflection to keep the theory conserving and thermodynamically consistent.  We split the self-energy to two. One with odd and the other with even symmetry with respect to the symmetry-breaking field of the critical point of the two-particle vertex. They will be related to the two-particle vertex differently in approximate schemes.

The odd self-energy stands for the order parameter emerging below the critical point with a diverging vertex. The system with the repulsive interaction is driven in intermediate coupling towards a magnetic order. The odd self-energy must then enter the Ward identity in order to keep thermodynamic consistency between the criticality in the two-particle vertex and the order parameter in the symmetry-broken phase. We argued in previous publications \cite{Janis:2017aa,Janis:2017ab,Janis:2019aa} that it is sufficient to obey the Ward identity only in the leading linear order in the symmetry-breaking (magnetic) field to describe the critical behavior qualitatively correctly. The odd self-energy determined from the static irreducible vertex $\Lambda$ satisfying the linearized Ward identity is
\begin{equation}\label{eq:WI-static}
\Delta\Sigma = -\frac{\Lambda}2 m\,.
\end{equation}         

There is no critical behavior in the charge sector and the even self-energy does not affect the critical behavior near the transition to the magnetic state. It  need not be related to the two-particle irreducible vertex via the Ward identity. It is responsible for the charge dynamics and should obey the Schwinger-Dyson equation of motion. Its mean-field (static) version is just the Hartree-Fock spin symmetric approximation. We then have
\begin{subequations}\label{eq:HF-normal}
\begin{equation}
\Sigma_{0}(\omega) = \frac U2 n \,.
\end{equation}   
Analogously the anomalous self energy, that has no odd part, is  
\begin{equation}
\mathcal{S}_{0}(\omega) = U \nu \,,
\end{equation}
\end{subequations}  
is proportional to the density of the Cooper pairs on the impurity. 
 
 The components determining the one-electron Green function of the superconducting quantum dot  in the mean-field approximation are   
\begin{subequations}
\begin{align}
X_{\sigma}(\omega) &= \omega\left[1 + s(\omega) \right] \nonumber 
\\
& \quad + \sigma \left(h + \frac{\Lambda}2m\right) - \left(\epsilon + \frac U2 n\right) \,, \\
\label{eq:Lambda-static}
Y(\omega) &= \Delta s(\omega) - U\nu \,.
\end{align}
\end{subequations}
The charge and spin densities are
\begin{subequations}
\begin{align} 
n &= \frac 1\beta\sum_{\omega_{n}}e^{i\omega_{n}0^{+}}\left[G_{\uparrow}(i\omega_{n}) + G_{\downarrow}(i\omega_{n})\right] \,,
\\
m &= \frac 1\beta\sum_{\omega_{n}}e^{i\omega_{n}0^{+}}\left[G_{\uparrow}(i\omega_{n}) - G_{\downarrow}(i\omega_{n})\right]\,,
\end{align} 
and the density of the Cooper pairs on the dot is
\begin{align} 
\nu c_{\Phi} &= \frac 1{2\beta}\sum_{\omega_{n}}e^{i\omega_{n}0^{+}}\left[\mathcal{G}_{\uparrow}(i\omega_{n}) + \mathcal{G}_{\downarrow}(i\omega_{n})\right] \,.
\end{align} 
\end{subequations}

The equations for the effective interaction $\Lambda$, the density of Cooper pairs $\nu$,  the charge density $n$, and magnetization $m$ close our mean-field approximation. It is free of the unphysical and spurious finite-temperature transition to the magnetic state due to the two-particle self-consistency. It qualitatively correctly describes the behavior of the quantum dot in weak as well as in strong coupling, including the Kondo regime for the dot attached to metallic leads. It can be applied at all temperatures and also in an arbitrary magnetic field. This mean-field approximation serves as the proper starting point for the perturbation expansion to include dynamical corrections. The mean-field  one-particle Green functions  replace the bare propagators in the perturbation expansion around the mean-field solution. We call them  thermodynamic propagators.   
 
 \subsection{Spectral representation}
 
The whole mean-field approximation can be fully solved in the Matsubara formalism.  What cannot be determined from the Matsubara frequencies are the spectral properties of the one and two-particle Green functions. To determine also the spectral properties one has to perform analytic continuation to the real frequencies. One needs to rewrite the sum over Matsubara frequencies to integrals with Fermi and Bose distribution functions. 

The  one-electron Green functions have a gap around the Fermi energy.  Since the hybridization self-energy $s(z)$ has a square-root singularity at the gap/band edges, the gap is fixed in the one-electron Green function and does not depend on the interaction strength. The poles and the band edges of the higher-order Green functions do, however, depend on the interaction strength. We hence must be careful when treating the two-particle functions in the spectral representation.
 
 The sum over the fermionic Matsubara frequencies for the one-particle function can then be rewritten in the spectral representation   
\begin{multline}\label{eq:AC-Fermi}
\frac{1}{\beta}\sum_n F(i\omega_n)e^{i\omega_{n}0^{+}}\rightarrow \ \sum_i f(\omega_i)\Res[F,\omega_i]
\\
-\ \left[\int_{-\infty}^{-\Delta} + \int_{\Delta}^{\infty}\right] \frac{d\omega}{\pi} f(\omega)\Im F(\omega+i0)  \,.
\end{multline}
Functions with bosonic symmetry have no gap in their spectra at non-zero temperatures with discrete Matsubara frequencies. 

The Andreev bound states are determined from zeros of the denominator $D_{\sigma}$ from Eq.~\eqref{eq:D-Sigma}. The frequencies of the poles of the one-electron Green function in the mean-field approximation are  
\begin{multline}\label{eq:omega-h}
\omega_{\sigma}(1+ s_{\sigma}) = - \sigma \left(h + \frac{\Lambda}2m\right) 
\\
+\ \sqrt{\left(\epsilon + \frac U2 n\right)^{2} + c_{\Phi}^{2}\left(s_{\sigma}\Delta - U\nu\right)^{2}} \,.
\end{multline}
The other two frequencies of the gap states are symmetrically situated on the other side of the Fermi energy.  

The spectral representation for the densities are
\begin{subequations}\label{eq:nmnu-thermo}
\begin{align}
n  &= n_{g} + n_{b} =\sum_{\alpha,\sigma} f(\alpha\sigma\omega_{\sigma})\Res[G_{\sigma},\alpha\sigma\omega_{\sigma}] \nonumber \\
&\qquad- \sum_{\sigma}\left[\int_{-\infty}^{-\Delta} + \int_{\Delta}^{\infty}\right] \frac{d\omega}{\pi} f(\omega)\Im G_{\sigma}(\omega_{+}) \,, 
\\
m &= m_{g} + m_{b} =  \sum_{\alpha,\sigma}\sigma f(\alpha\sigma\omega_{\sigma})\Res[G_{\sigma},\alpha\sigma\omega_{\sigma}] \nonumber \\
&- \sum_{\sigma}\sigma\left[\int_{-\infty}^{-\Delta} + \int_{\Delta}^{\infty}\right] \frac{d\omega}{\pi} f(\omega) \Im G_{\sigma}(\omega_{+})\,,
\\
c_{\Phi} \nu   &= c_{\Phi}\left(\nu_{g} + \nu_{b}\right) = \frac 12\sum_{\alpha,\sigma} f(\alpha\sigma\omega_{\sigma})\Res[\mathcal{G}_{\sigma},\alpha\sigma\omega_{\sigma}] \nonumber \\
&\  - \frac 12\sum_{\sigma}\left[\int_{-\infty}^{-\Delta} + \int_{\Delta}^{\infty}\right] \frac{d\omega}{\pi} f(\omega)\Im \mathcal{G}_{\sigma}(\omega_{+}) \,.
\end{align}
\end{subequations}
We abbreviated the notation of the frequency with an infinitesimal imaginary part  $\omega + i0^{+} = \omega_{+}$. We split the contributions to the densities to those from the in-gap states, subscript $g$ and from the band states, subscript $b$. Notice that the density of the Cooper pairs from the gap and band states is now spin dependent.  

The residues of the one-electron Green function are
\begin{subequations}
 \begin{align}
 \Res\left[G_{\sigma},\sigma\sigma'\omega_{\sigma'}\right] &= \frac{1}{K_{\sigma'}}\left[X_{\sigma'} + \sigma\sigma' \left(\epsilon + \frac U2 n\right)\right] \,,\\
 \Res\left[\mathcal{G}_{\sigma},\sigma\sigma'\omega_{\sigma'}\right] &= -\frac {\sigma' c_{\Phi}}{ K_{\sigma'}}\left[s_{\sigma'}\Delta  -U\nu\right] \,,
 \end{align}
 \end{subequations}
with
\begin{subequations}
 \begin{align}
X_{\sigma} &= \sqrt{\left(\epsilon + \frac U2 n\right)^{2} + c_{\Phi}^{2} \left(s_{\sigma}\Delta  - U\nu\right)^{2}}  \,,\\
K_{\sigma} &=  2X_{\sigma}\left[1 + \frac{\Delta^{2}s_{\sigma}}{\Delta^{2} - \omega_{\sigma}^{2}} \right]
 \nonumber\\ 
& \qquad - 2c_{\Phi}^{2}\left(s_{\sigma}\Delta - U\nu\right)\frac{\omega_{\sigma} s_{\sigma}\Delta}{\Delta^{2} - \omega_{\sigma}^{2}} \,.
 \end{align}
 \end{subequations}

The analytic representation of the two-particle Green and vertex functions is more complex. 
The integrand of the screening integral has no gap at non-zero temperatures with a simple analytic representation of the sum over bosonic Matsubara frequencies
\begin{multline}
\mathcal{X} = - P\int_{-\infty}^{\infty}\frac {dx}{\pi}b(x) \Im\left[\frac{\psi(x_{+})\psi(-x_{+})\phi(-x_{+})}{1 + \Lambda\phi(-x_{+})}\right]\,.
\end{multline}
The explicit analytic representations separating the gap and band contributions of the electron-hole $\phi_{\sigma}(\omega_{+})$ and  electron-electron $\psi(\omega_{+})$ are presented in Appendices A and B.

 \subsection{Full Green function and the spectral self-energy}

The spectral representation is necessary not only to determine the positions of the in-gap states. It is generally needed to disclose the whole spectral structure of the interacting system when we go beyond the mean-field approximation in the perturbation expansion. The first step beyond the static theory are dynamical corrections to the static self-energy.  The even self-energy is determined from the dynamical Schwinger-Dyson equation of motion. Its form with the static irreducible vertex $\Lambda$  is for the normal part  
\begin{subequations}\label{eq:SD-full}
\begin{multline}
\Sigma^{Sp}(\omega_{+}) = - U \int_{-\infty}^{\infty} \frac{d x}{\pi} \left\{f(x) \frac{\Im \bar{G}^{Sp}(x_{+})}{1 + \Lambda \phi(x - \omega_{+})}  
\right. \\ \left.
-\ b(x) \bar{G}^{Sp}(\omega_{+} + x) \Im\left[\frac{1}{1 + \Lambda\phi(x_{+})} \right]\right\} \,
\end{multline}
and analogously for the anomalous self-energy
\begin{multline}
c_{\Phi}\mathcal{S}^{Sp}(\omega_{+}) =- U \int_{-\infty}^{\infty} \frac{d x}{\pi} \left\{f(x) \frac{\Im \bar{\mathcal{G}}^{Sp}(x_{+})}{1 + \Lambda \phi(x - \omega_{+})}  
\right. \\ \left.
 -\ b(x) \bar{\mathcal{G}}^{Sp}(\omega_{+} + x) \Im\left[\frac{1}{1 + \Lambda\phi(x_{+})} \right]\right\} \,.
\end{multline}
\end{subequations}
The integrand in the Schwinger-Dyson equation contains two parts, the two-particle and one-particle ones. The former part, consisting of the electron-hole bubble $\phi(\omega_{+})$ and vertex $\Lambda$, controls the thermodynamic response and the critical behavior. It hence must be the same as used to determine the two-particle irreducible vertex $\Lambda$ and the odd self-energy $\Delta\Sigma$. The one-particle propagators $G^{Sp}(\omega_{+})$ and $\mathcal{G}^{Sp}(\omega_{+})$ in the Schwinger-Dyson equation carry information about the spectral properties. Its odd self-energy $\Delta\Sigma$ must be identical with that used to determine the two-particle vertex. Its noncritical even self-energy $\Sigma^{Sp}(\omega_{+})$ can be selected self-consistently containing the spectral self-energy, a solution of the Schwinger-Dyson equation. Since the Schwinger-Dyson equation determines only the spin-symmetric self-energy we used  the spin-averaged propagators $\bar{G}^{Sp}(x_{+}) =\left(G^{Sp}_{\uparrow}(x_{+}) + G^{Sp}_{\downarrow}(x_{+})\right)/2$ and $\bar{\mathcal{G}}^{Sp}(x_{+}) =\left(\mathcal{G}^{Sp}_{\uparrow}(x_{+}) + \mathcal{G}^{Sp}_{\downarrow}(x_{+})\right)/2$. 

The one-particle propagators $G_{\sigma}^{Sp}$ and $\mathcal{G}_{\sigma}^{Sp}$ used in the Schwinger-Dyson equation then are
\begin{subequations}
\begin{align}
G^{Sp}_{\sigma}(\omega_{+} ) &=  \frac{\omega + \sigma\left(h -\Delta\Sigma\right) + \epsilon  + \Sigma^{Sp}(-\omega_{+})}{D^{Sp}_{\sigma}(\omega_{+})} 
\,, \\
\mathbb{\mathcal{G}}^{Sp}_{\sigma}(\omega_{+}) &= -c_{\Phi} \frac{s(\omega_{+})\Delta  - \mathcal{S}^{Sp}(\omega_{+})}{D^{Sp}_{\sigma}(\omega_{+})}
\end{align}
\end{subequations}
with the denominator
\begin{multline}
D^{Sp}_{\sigma}(\omega_{+}) = \left[\omega_{+} + \sigma\left(h -\Delta\Sigma\right)  - \epsilon  - \Sigma^{Sp}(\omega_{+})\right]
 \\ 
\times \left[\omega_{+}  + \sigma\left(h -\Delta\Sigma\right)  + \epsilon + \Sigma^{Sp}(-\omega_{+})\right]  
\\
 -\ c_{\Phi}^{2}\left[s(\omega_{+})\Delta - \mathcal{S}^{Sp}(\omega_{+})\right]^{2}  \,,
\end{multline}
where $\Delta\Sigma = -\Lambda m^{T}/2$ and $m^{T}$ is the magnetization calculated with the thermodynamic propagator determining the effective interaction $\Lambda$. 

The normal  dynamical self-energy $\Sigma^{Sp}(\omega_{+})$ and the anomalous one  $\mathcal{S}^{Sp}(\omega_{+})$  from the Schwinger-Dyson equation~\eqref{eq:SD-full} and the odd one $\Delta\Sigma$ from  the Ward identity, Eq.~\eqref{eq:WI-static}  are the physical self-energies. It means that a mean-field approximation at the two-particle level generates nontrivial dynamical contributions to the one-particle self-energy in an analogous manner as the random-phase approximation generates a dynamical self-energy for the Hartree-Fock mean-field thermodynamics.  We will analyze the dynamical corrections from the Schwinger-Dyson equation  in a separate paper. 

\section{Gap states and $0-\pi$ transition}
 
The spectral representation is needed for the determination of the positions of the in-gap states and finding the point of their crossing signaling the $0-\pi$ transition at zero temperature. We need to keep the applied magnetic field positive in order to be able to continue the solution from the weak-coupling $0$-phase to the strong-coupling $\pi$-phase. We resort to the static mean-field approximation to determine the $0-\pi$ transition.
 
 We split the contributions from the band and gap states and introduce the following abbreviations
 \begin{subequations}
 \begin{align}
 \epsilon_{U}& = \epsilon + \frac U2 n \,,
 \\
 \Gamma_{\sigma}&= s_{\sigma}\Delta - U\nu\,.
 \end{align}
 \end{subequations}
 
 The one-electron parameters are
 \begin{align}
n - n_{b} = n_{g} &=  \frac{1}{K_{\uparrow}K_{\downarrow}}\sum_{\sigma}K_{\sigma}\left[X_{\bar{\sigma}} - \epsilon_{U}\Delta f_{\bar{\sigma}}\right]  \,, \\ 
m - m_{b} = m_{g} &=  \frac 1{K_{\uparrow}K_{\downarrow}}\sum_{\sigma}K_{\sigma}\left[X_{\bar{\sigma}}\Delta f_{\bar{\sigma}} - \epsilon_{U}\right]  \,, \\ 
 \nu - \nu_{b} = \nu_{g} &=  \frac{1}{2K_{\uparrow}K_{\downarrow}}\sum_{\sigma}K_{\sigma}\Gamma_{\bar{\sigma}}\Delta f_{\bar{\sigma}}\,, 
 \end{align}
where the subscripts $b,g$ refer to the band and gap contributions, respectively. We denoted $\Delta f_{\sigma} = f(-\omega_{\sigma}) - f(\omega_{\sigma})$.   We recall that the poles of the mean-field propagators $G_{\sigma}(\omega)$ and $\mathcal{G}_{\sigma}(\omega)$ are $\omega_{\sigma}$ and $-\omega_{\bar{\sigma}}$. The equations for the in-gap-state frequencies are determined in Eq.~\eqref{eq:omega-h}.

The $0-\pi$ transition in the external magnetic field in a spin-polarized state happens at $\omega_{\uparrow} = 0$, that is   
\begin{equation}\label{eq:CriticalU-magnetic}
h + \frac \Lambda2 m  = \sqrt{\left(\epsilon + \frac U2 n\right)^{2} + c_{\Phi}^{2}\left(s_{\sigma}\Delta - U\nu\right)^{2}}\,.
\end{equation}
This equation tells us that the effective interaction $\Lambda$ affects the transition only in the spin-polarized solution with $h>0$. The transition in our mean-field approximation with no spectral self-energy at $h=0$ coincides with the Hartree-Fock result. It is, however, important to realize that unlike the Hartree-Fock solution the mean-field approximation with an effective interaction $\Lambda$ is free of the spurious transition to the magnetic state at non-zero temperatures and is thermodynamically consistent in the whole range of the input parameters. Notice, however, that the effective interaction does affect the position of the $0-\pi$ transition in the spin-symmetric state if we employ the spectral self-energy from Eq.~\eqref{eq:SD-full}.

\subsection{Spin-symmetric state}

We first approach the $0-\pi$  transition from the weak-coupling regime in the spin-symmetric state. We then have $X_{0} = \sqrt{\epsilon_{U}^{2} + c_{\Phi}^{2}\left(s_{0}\Delta - U\nu\right)^{2}}$ with $s_{0}= \Gamma/(\Delta^{2} - \omega_{0}^{2})$ and $\omega_{0}(1 + s_{0}) = X_{0}$. Further on,
\begin{multline}
K_{0} = 2\kappa_{0} X_{0} = 2\left\{1 + \frac{s_{0}\Delta}{(1 + s_{0})\left(\Delta^{2} - \omega_{0}^{2}\right)}
\right. \\ \left.
\times\left[\Delta + c_{\Phi}^{2}U\nu + s_{0}\Delta\left(1 - c_{\Phi}^{2}\right) \right] \right\}X_{0} \,.
\end{multline}
The equation for the positive frequency of the gap state is
\begin{multline}\label{eq:Omega0-symmetric}
\left[(1 + s_{0})\kappa_{0}\omega_{0} + \frac U2 \tanh\left(\frac{\beta\omega_{0}}2 \right)\right]^{2}
\\
 = \left(\kappa_{0}\epsilon_{b} + \frac U2 \right)^{2} + c_{\Phi}^{2}\kappa_{0}^{2} \Gamma_{0b}^{2} \,, 
\end{multline}
where we used an identity $\Delta f(\omega) = \tanh\left(\beta\omega/2 \right)$.
We denoted $\epsilon_{b} = \epsilon + Un_{b}/2$ and $\Gamma_{0b} = s(\omega_{0})\Delta - U\nu_{b}$. There is aways a solution for $\omega_{0}>0$ for arbitrary $U$ at non-zero temperature. There is hence no crossing of the in-gap states at non-zero temperature in the spin-symmetric state as already observed in Ref.~\cite{Janis:2016aa}. 

The in-gap-state frequency  reaches the Fermi energy, that is $\omega_{0} = 0$, at the $0-\pi$ transition only at zero temperature.   The spin-symmetric state can reach the critical interaction strength $U_{c}$ of the $0-\pi$ transition only for $\beta\omega_{0} = \infty$ defining a quantum critical point. The equation for the critical interaction  reads  
\begin{equation}\label{eq:Uc-symmetric-MF}
\frac{U^{2}_{c}}4 = \left[\left(1 + \frac{\Gamma}\Delta \right)\epsilon_{b} + \frac {U_{c}}2 \right]^{2} + c_{\Phi}^{2}\left(1 + \frac{\Gamma}\Delta \right)^{2} \Gamma_{0b}^{2} \,. 
\end{equation}

The equilibrium spin-symmetric solution must be stable with respect to the perturbations caused by a small magnetic field.  Its local stability is determined from the static magnetic susceptibility. It is critically dependent on the effective interaction of the mean-field approximation.  The mean-field static susceptibility has the Stoner form
\begin{equation}\label{eq:MF-susceptibility}
\chi = - \frac{2\phi(0)}{1 + \Lambda\phi(0)} \,.
\end{equation}
The denominator on the right-hand side of Eq.~\eqref{eq:MF-susceptibility} is positive at any temperature and non-diverging at non-zero temperatures due to the appropriately chosen screening of the interaction strength in the self-consistent equation~\eqref{eq:Lambda-general}.  The susceptibility can diverge only at zero temperature in the $\pi$-phase as we demonstrate later.

\subsection{Magnetic state}

We introduce an effective magnetic field containing the entire effect of the applied magnetic field in the mean-field approximation to simplify the notation
\begin{equation}
h_{\Lambda} =h\left(1 + \frac{\Lambda m}{2h}\right) \,.
\end{equation}

 The crossing of the in-gap states takes place when $\omega_{\uparrow}= 0$ for which $\omega_{\downarrow}(1 + s_{\downarrow}) = 2h_{\Lambda}$.  Consequently, $K_{\uparrow}= 4h_{\Lambda}(1 + \Gamma/\Delta)$, $X_{\uparrow}^{2} = \epsilon_{U}^{2} + c_{\Phi}^{2}(\Gamma - U\nu)^{2}$ and 
\begin{subequations}  
\begin{align}
X_{\downarrow}^{2} &= \epsilon_{U}^{2} + c_{\Phi}^{2} \left(\frac{\Gamma\Delta}{\sqrt{\Delta^{2} - \omega_{\downarrow}^{2}}} - U\nu\right)^{2} \,,
\\
K_{\downarrow} &=4 h_{\Lambda}\left\{1 + \frac{\Gamma\Delta}{\left(\Delta^{2} - \omega_{\downarrow}^{2} \right)^{3/2}}\nonumber 
\right.\\ 
&\left. \qquad \times\left[\Delta - c_{\Phi}^{2}\left(\frac{\Gamma\Delta}{\sqrt{\Delta^{2} - \omega_{\downarrow}^{2}}} - U\nu \right)\right] \right\}
\end{align}
\end{subequations}

We further have $\Delta f_{\uparrow} = 0$ at the crossing point at non-zero temperatures  and hence
\begin{subequations}  
\begin{equation}
\epsilon_{U} = \frac{U\left[K_{\downarrow}\Delta + 2X_{\downarrow}(\Gamma + \Delta)\right] + 4X_{\downarrow}(\Gamma + \Delta)\epsilon_{b} }{(\Gamma + \Delta)\left(2K_{\downarrow} + U\Delta f_{\downarrow}\right)} \,,
\end{equation}
\begin{equation}
\Gamma - U\nu = \frac{2K_{\downarrow}\Gamma_{b} + U\Delta f_{\downarrow}\left(\Gamma - s_{\downarrow}\Delta \right)}{2K_{\downarrow} + U\Delta f_{\downarrow}} \,.
\end{equation}
\end{subequations}

The equation for  frequency $\omega_{\downarrow}$  at the crossing is
\begin{multline}\label{eq:0-pi-transition-magnetic}
\Delta^{2}\left(2K_{\downarrow} + U\tanh\left(\frac{\beta\omega_{\downarrow}}2 \right)\right)^{2} 
\\
= \left[UK_{\downarrow} + 4X_{\downarrow}\left(1 + \frac{\Gamma}{\Delta}\right)\left(\epsilon_{b} + \frac U2  \right)\right]^{2} 
\\
+\ c_{\Phi}^{2}\left[2K_{\downarrow}\Gamma_{b} + U\tanh\left(\frac{\beta\omega_{\downarrow}}2 \right)\left(\Gamma - s_{\downarrow}\Delta \right)\right]^{2} \,.
\end{multline}
The crossing leads to the $0$-$\pi$ transition only at zero temperature and zero magnetic field and it is a quantum critical point with a diverging magnetic susceptibility when approached from the spin-symmetric state.    

The solution of Eq.~\eqref{eq:0-pi-transition-magnetic} shows a universal behavior for non-zero magnetic field. We can divide all energy variables $T,U,\Lambda,\Delta,\epsilon, K, X,\Gamma$, by $2h>0$ to turn them dimensionless.  The dimensionless solution for $\bar{\omega}_{\downarrow} = 1 + \bar{\Lambda} m$ will then become universal, independent of the actual value of the applied magnetic field. 

\section{Asymptotic atomic limit}

The important test of reliability of the approximations is a comparison with the existing exact solutions in specific limiting situations. The present two-particle approximation with the effective interaction $\Lambda$ from Eq.~\eqref{eq:Lambda-general} and the spectral self-energy from Eq.~\eqref{eq:SD-full} was shown to reproduce qualitatively correctly the Kondo regime of the exact solution of the SIAM for $\Delta=0$. The opposite asymptotic atomic limit $\Delta\to\infty$ can also be exactly solved \cite{Rozhkov:2000aa,Vecino:2003aa,Bauer07,Meng:2009aa,Meden:2019aa}. Here we compare the predictions of the presented mean-field approximation with the exact results of the atomic limit with no hybridization to the band electrons. The exact results of the atomic limit are summarized in Appendix C. 

The normal and anomalous Green functions in the atomic limit are
\begin{subequations}
\begin{align}
G_{\sigma}(\omega) &= \frac 1{2X_{0}}\left[\frac{X_{0} + \epsilon_{U}}{\omega - \omega_{\sigma}} +  \frac{X_{0} - \epsilon_{U}}{\omega + \omega_{\bar{\sigma}}}\right]
\\
\mathcal{G}_{\sigma}(\omega) &= - \frac{c_{\Phi}\Gamma}{2X_{0}}\left[\frac 1{\omega - \omega_{\sigma}} - \frac 1{\omega + \omega_{\bar{\sigma}}} \right]\,.
\end{align}
\end{subequations}  
where $\bar{\sigma}= - \sigma$ and 
\begin{subequations}
\begin{align}
\omega_{\sigma} &= \bar{\sigma} h_{\Lambda} + X_{0} \,,
\\
X_{0} &= \sqrt{\epsilon_{U}^{2} + c_{\Phi}^{2}\left(\Gamma - U\nu \right)^{2}} \,.
\end{align}
\end{subequations}

The explicit value of the electron-hole bubble is
\begin{equation}
\phi(\omega_{+}) =\frac{\phi_{\uparrow\downarrow}\Delta\omega}{\omega_{+}^{2} - \Delta\omega^{2}} \,,
\end{equation}
where we denoted $\phi_{\uparrow\downarrow} = (\Delta f_{\downarrow} - \Delta f_{\uparrow})/2 = - \phi(0)\Delta\omega = m$ and $\Delta\omega = \omega_{\downarrow} - \omega_{\uparrow}$.  Further on,
\begin{equation}
\frac{\phi(\omega_{+})}{1 + \Lambda\phi(\omega_{+})} = \frac{\phi_{\uparrow\downarrow}\Delta\omega}{\omega_{+}^{2} - \omega_{\phi}^{2}} \,,
\end{equation} 
where the poles $\pm \omega_{\phi}$ of this function are  
\be
\omega_{\phi}^{2} = \Delta\omega\left[\Delta\omega - \Lambda \phi_{\uparrow\downarrow} \right] = \Delta\omega^{2}\left[1 + \Lambda \phi(0) \right] \,.
\ee 
The particle-particle bubble in the atomic limit is 
%
\begin{multline}
\psi(\omega) 
\\
= \frac {\left(\Delta f_{\uparrow} + \Delta f_{\downarrow}\right)}{8X_{0}^{2}}\left[\frac{\left(X_{0} - \epsilon_{U} \right)^{2}}{\omega + 2X_{0}} - \frac{\left(X_{0} + \epsilon_{U} \right)^{2}}{\omega - 2X_{0}}\right] \,.
\end{multline}
We further use the following identities
\begin{subequations}
\begin{align}
\frac 1{1 + \Lambda \phi(0)} &= 1 + \frac{\Lambda m}{2h} \,, \\
\Delta\omega & = 2h \left(1 + \frac{\Lambda m}{2h}\right) \, \\
\omega_{\phi} &= 2h\sqrt{1 + \frac{\Lambda m}{2h}} \,
\end{align}
\end{subequations}
to represent the screening integral
\begin{widetext}
\begin{multline}
\mathcal{X} = \frac{\left(\Delta f_{\downarrow} + \Delta f_{\uparrow} \right)^{2}m}{64 X_{0}^{4}\left(4X_{0}^{2} - \omega_{\phi}^{2}\right)} \sqrt{1 + \frac{\Lambda m}{2h}}\left\{ \frac{\left(X_{0} + \epsilon_{U} \right)^{4} + \left(X_{0} - \epsilon_{U} \right)^{4}}{4X_{0}}\left[2X_{0}\coth\left(\frac{\beta\omega_{\phi}}2 \right) - \omega_{\phi}\coth(\beta X_{0})\right] 
\right. \\ \left.
+\ \frac{\left(X_{0}^{2} - \epsilon_{U}^{2}\right)^{2}}{\left(4X_{0}^{2} - \omega_{\phi}^{2}\right)}\left[\left(4X_{0}^{2} + \omega_{\phi}^{2}\right)\coth\left(\frac{\beta\omega_{\phi}}2 \right) - 4X_{0}\omega_{\phi}\coth(\beta X_{0})
 - \frac{\beta\omega_{\phi}\left(4X_{0}^{2} - \omega_{\phi}^{2}\right)}{2\sinh^{2}(\beta X_{0})} \right]\right\}  \,.
\end{multline}
We used $\Delta b(\omega) = b(\omega) - b(-\omega)= \coth(\beta\omega/2)$. 
The screening integral at half filling, $\epsilon_{U}= 0$, reduces to
\begin{equation}
\mathcal{X} = \frac{\left(\Delta f_{\downarrow} + \Delta f_{\uparrow} \right)^{2}m}{256 X_{0}^{3}}\sqrt{1 + \frac{\Lambda m}{2h}}\left[ 2X_{0}\coth\left(\beta h\sqrt{1 + \frac{\Lambda m}{2h}}\right) - 2h\sqrt{1 + \frac{\Lambda m}{2h}}\ \frac{\cosh\left(\beta X_{0} \right) + \beta X_{0}}{2\sinh^{2}(\beta X_{0})}\right]  
\end{equation}
\end{widetext}
and in the spin-symmetric state, $\omega_{\uparrow}\nearrow\omega_{\downarrow} = X_{0}$ to
\begin{multline}
\mathcal{X} = \frac{T\chi}{128 X_{0}^{6}}\tanh^{2}\left(\frac{\beta X_{0}}{2}\right)\left[\left(X_{0} + \epsilon_{U} \right)^{4}  
\right. \\ \left.
+\  \left(X_{0} - \epsilon_{U} \right)^{4} + 2\left(X_{0}^{2} - \epsilon_{U}^{2}\right)^{2}\right] \,,
\end{multline}
where 
\begin{equation}\label{eq:susc-paramagnetic}
\chi = \lim_{h\to0}\frac m h = \frac{2f(X_{0})(1 - f(X_{0})) }{T - 2\Lambda f(X_{0})(1 - f(X_{0})) }  
\end{equation}
 is the magnetic susceptibility. 

The spin symmetric solution is locally stable at all non-zero temperatures. If we introduce a generalized Kondo scale $a= 1 +\Lambda\phi(0)$ and assume that the solution approaches the critical point $a \searrow 0$ then the asymptotic critical solution is   
\begin{subequations}\label{eq:Lambda-critical} 
\begin{align}
\Lambda &= \frac T{2f_{0}(1 - f_{0})} \,,
\\
a&= \frac{ T^{3}C}{128 U X_{0}^{2} n^{2}f_{0}^{2}(1 - f_{0})^{2}}\,
\end{align}
and 
\begin{multline}
C = \frac{1}{X_{0}^{4}}\tanh^{2}\left(\frac{\beta X_{0}}{2}\right)\left[\left(X_{0} + \epsilon_{U} \right)^{4}  
\right. \\ \left.
+\  \left(X_{0} - \epsilon_{U} \right)^{4} + 2\left(X_{0}^{2} - \epsilon_{U}^{2}\right)^{2}\right] \,.
\end{multline}
\end{subequations}
The magnetic susceptibility can diverge only at zero temperature. The critical region can, however, be reached only if the product $f_{0}(1 - f_{0}) \ge T/U$ with the decreasing temperature so that $U\ge \Lambda >0$. 

The spin-symmetric solution is  identical with the Hartree-Fock one. It becomes  exact at zero temperature.  The boundary for the $0$-phase at zero temperature form Eq.~\eqref{eq:Uc-symmetric-MF} is
\begin{equation}
\frac{U_{c}^{2}}{4} = \left(\epsilon + \frac U2 \right)^{2} + c_{\Phi}^{2}\Gamma^{2} \,,
\end{equation} 
which is the exact result for the $0-\pi$ transition in the atomic limit \cite{Vecino:2003aa,Meng:2009aa}.

Resolving the particle and Cooper-pair densities in the spin-polarized state we obtain
\begin{subequations}\label{eq:n-nu-thermo}
\begin{align}
n &= 2\frac{2X_{0} - \epsilon \left(\Delta f_{\downarrow} + \Delta f_{\uparrow} \right)}{4X_{0} +  U\left(\Delta f_{\downarrow} + \Delta f_{\uparrow}\right)}  \,,\\
\nu &= \frac{ \Gamma \left(\Delta f_{\downarrow} + \Delta f_{\uparrow} \right)}{4X_{0} +  U \left(\Delta f_{\downarrow} + \Delta f_{\uparrow}\right)}\,.
\end{align}
\end{subequations}
\begin{subequations}
\begin{align}
\epsilon_{U} &= \frac{4X_{0}\delta}{4X_{0} + U\left(\Delta f_{\uparrow} + \Delta f_{\downarrow}\right)} 
 \nonumber\\
&= \delta\left[1 - \frac U{2U_{c}} \left(\Delta f_{\uparrow} + \Delta f_{\downarrow}\right)\right]\,,\\
\Gamma - U\nu &= \frac{4X_{0}\Gamma}{4X_{0} + U\left(\Delta f_{\uparrow} + \Delta f_{\downarrow}\right)}  \,,
\end{align}
\end{subequations}
where we denoted $\delta = \epsilon + U/2$ and $U_{c} = 2\sqrt{\delta^{2} + c_{\Phi}^{2}\Gamma^{2}}$. 

The equation for  $X_{0}$ needed to obtain $n$ and $\nu$ is 
 \begin{equation}\label{eq:X0-general}
\left[2X_{0} + \frac U2\left(\Delta f_{\uparrow} + \Delta f_{\downarrow}\right)\right]^{2}  = 4\left(\delta^{2} + c_{\Phi}^{2}\Gamma^{2}\right) = U_{c}^{2} \,.
\end{equation}
Since $X_{0}\ge 0$ then  
\begin{align}\label{eq:X0-U}
2X_{0} &= U_{c} - \frac U2\left(\Delta f_{\uparrow} + \Delta f_{\downarrow}\right)\ge 0\,.
\end{align}

Using this solution we obtain
\begin{subequations}\label{eq:nnu-def}
\begin{align}
n &= 1 - \frac{\delta}{U_{c}}\left(\Delta f_{\uparrow} + \Delta f_{\downarrow}\right)\,,\\
\nu &= \frac{\Gamma\left(\Delta f_{\uparrow} + \Delta f_{\downarrow}\right)}{2U_{c}}\,,
\end{align}
\end{subequations}
and 
\begin{align} \label{eq:m-def}
m &= \frac 12 \left(\Delta f_{\downarrow} - \Delta f_{\uparrow} \right) \,.
\end{align}

Since $\omega_{\uparrow} = -(h + \Lambda m/2) + X_{0}$ and  $\omega_{\downarrow} = (h + \Lambda m/2) + X_{0}$,  we have always three independent variables to determine self-consistently, $\Lambda, X_{0}$, and $m$. The three coupled equations determining these variables are  Eq.~\eqref{eq:Lambda-cubic}, Eq.~\eqref{eq:X0-U},  and Eq.~\eqref{eq:m-def}. The charge density and the density of the Cooper pairs are then calculated from Eqs.~\eqref{eq:nnu-def}. 

The equation determining the crossing of the in-gap states  in the applied magnetic field at non-zero temperature is $\omega_{\uparrow}= 0$ and $\Delta f_{\uparrow}= 0$, hence 
\begin{equation}
U_{c} - \frac U2\tanh\left(\frac{\beta\omega_{\downarrow}}2 \right) = 2h\left(1 + \frac{\Lambda m}{2h} \right) = \omega_{\downarrow}\,.
\end{equation}
There is always a crossing of the in-gap states  for arbitrary interaction $U$ at non-zero temperature at an appropriate magnetic field. 

The zero-temperature solution behaves differently. An infinitesimally small magnetic perturbation generates a fully polarized magnetic state in the $\pi$-phase, $U> U_{c}$. The $\pi$-phase is bounded by $\omega_{\uparrow}<0$, where $m=1$ from Eq.~\eqref{eq:m-def}, $n=1$, and $\nu=0$ from Eqs.~\eqref{eq:nnu-def} for $\beta=\infty$.   The effective interaction $\Lambda=U$ at zero temperature and the screening integral from Eq.~\eqref{eq:Lambda-general} is proportional to $n_{\uparrow}n_{\downarrow}= (n - m)^{2}/4$ in the $\pi$-phase. It then means that $\omega_{\downarrow} = U_{c} + U$ while $\omega_{\uparrow} = U_{c} - U$ with $h\searrow 0$. The Hartree-Fock solution at zero temperature is exact also in the $\pi$-phase at zero temperature.  

There is a fundamental difference between the $0$-phase and the $\pi$-phase and we can distinguish the two phases by the low-temperature asymptotics of the magnetic susceptibility. We have $2X_{0}= U_{c} -U$ in the leading low-temperature asymptotics  in the $0$-phase, that is, for $U<U_{c}$ . Consequently, the magnetic susceptibility from Eq.~\eqref{eq:susc-paramagnetic} at low temperatures is
\begin{equation}
\chi \doteq \frac{8}T e^{-\beta(U_{c} - U)} \,,
\end{equation}   
which reflects the Meissner effect due to the presence of the singlet Bound states (ABS).

The situation in the $\pi$-phase, $U>U_{c}$, is quite different. Equation~\eqref{eq:X0-U} leads at low temperatures to
\begin{equation}
\beta X_{0} = \arctan\left(\frac{U_{c}}U \right) \left[ 1 - \frac{4TU}{U^{2} - U_{c}^{2}}\right] \,.
\end{equation}
The magnetic susceptibility is
\begin{align}
\chi &= \frac{U^{2} - U_{c}^{2}}{2 U^{2}} \ \frac 1a
\end{align}
with the effective interaction and the Kondo scale from Eqs.~\eqref{eq:Lambda-critical} are
\begin{align}
\Lambda &= \frac{2UT}{U^{2} - U_{c}^{2}} \,,
\\
a &= \frac{UU_{c}^{2}}{2\left(U^{2} - U_{c}^{2}\right)^{2}}  \arctan^{-2}\left(\frac{U_{c}}U \right) T \,.
\end{align}
The magnetic susceptibility follows the Curie law due to the presence of the local magnetic moment of the fermionic excitations on the dot.  The non-universal Curie constant 
\begin{equation}
C= \frac{\left(U^{2} - U_{c}^{2}\right)^{3}}{2U^{3}U_{c}^{2}}\arctan^{2}\left(\frac{U_{c}}U \right) 
\end{equation} 
is in the static mean-field approximation overestimated and grows with the increasing interaction strength. It lies above the exact value and will be corrected by the dynamical spectral self-energy. 

The observed behavior of the in the strong-coupling limit for $U>U_{c}$ discloses another feature of the solution in the external magnetic filed. We proved that the limit to zero magnetic field and to zero temperature do not commute. If we keep the magnetic field non-zero and limit the temperature to zero the solution behaves analytically and continuously reaches the fully saturated state at zero temperature.  On the other hand, if we keep the temperature non-zero and switch off the magnetic field we stay in the spin-symmetric states down to zero temperature where the magnetic susceptibility diverges and an infinitesimally small magnetic field lifts the degeneracy to a magnetically saturated state.     

\section{Numerical results}

We apply the mean-field approximation in the atomic limit to show its similarities and stress the substantial differences to the Hartree-Fock solution and to test the reliability of our mean-field approximation in different regimes. The major asset of the mean-field approximations is that they can be used in the whole range of the model parameters. They are qualitatively reliable if they do not lead to an unphysical and spurious behavior.  We know that the reduced parquet equations with a two-particle self-consistency reproduce qualitatively correctly the limit of the zero gap. It is instructive to apply it in the opposite limit of the infinite superconducting gap which is the least fitting situation for the application of the many-body construction. Many of the qualitative features of the solution in the atomic limit are generic and mimic the behavior of the finite-gap model except for the Kondo limit of the vanishing gap.

\subsection{Spin-symmetric solution}  

Our static mean-field solution in the spin-symmetric state, that is in the absence of the magnetic field, coincides with the Hartree-Fock approximation. This may seem a limiting factor, but it holds only at the one-particle level when the dynamical corrections in the spectral self-energy from the Schwinger-Dyson equation are neglected.  The first, and the most important difference between our mean-field theory and the Hartree-Fock approximation in the spin-symmetric state is the stability with respect to small fluctuations of the magnetic field, which is reflected in the magnetic susceptibility. 
\begin{figure}
\includegraphics[width=8.5cm]{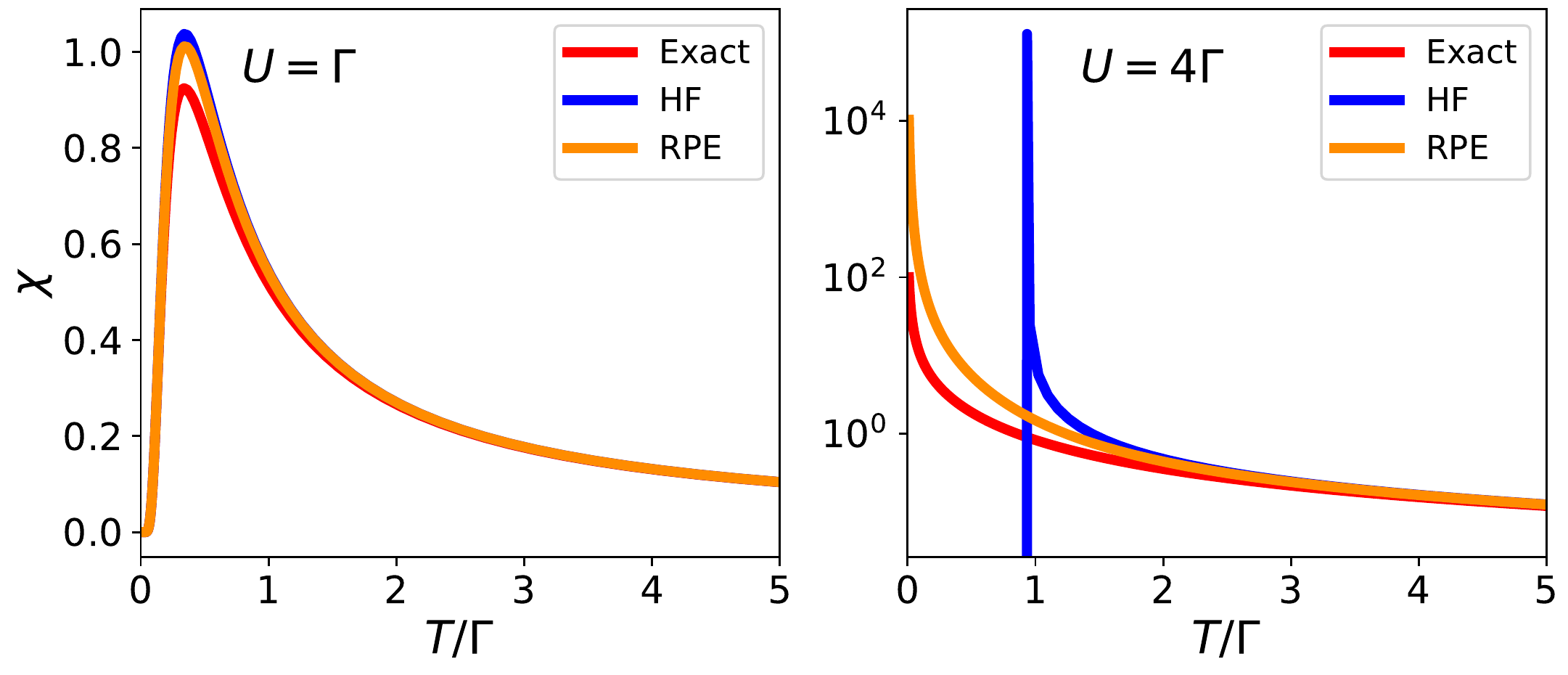}
\caption{Magnetic susceptibility as a function of temperature in the spin-symmetric state at half-filling in the $0$- phase ($U=\Gamma$) and the $\pi$-phase ($U=4\Gamma$) for the phase difference $\Phi=0$. Hartree-Fock (HF), reduced parquet equations (RPE) and exact (EXACT) solutions  are compared. The unphysical instability with the diverging susceptibility makes the Hartree-Fock mean-field solution in strong coupling unreliable at low temperature@`s.   \label{fig:Susc-temp} }
\end{figure}       

We plotted the magnetic susceptibility of the spin-symmetric state at half filling as a function of temperature in Fig.~\ref{fig:Susc-temp}. We compared the two mean-field approximations, our, based on the reduced parquet equations (RPE), and the Hartree-Fock one (HF), with the exact solution in the atomic limit. There is no big difference in the $0$-phase where all solutions asymptotically approach zero at zero temperature. Quite a different behavior is, however, observed in the $\pi$-phase. Both the exact and RPE solutions lead to a divergent susceptibility at zero temperature, while the HF solution predicts  an unphysical critical point with diverging susceptibility at a temperature of order of the hybridization strength $\Gamma$ that is taken as the energy unit. The magnetic susceptibility is a physical, measurable quantity being able to distinguish the character of the in-gap states. The in-gap states in the $0$-phase are bound pairs, ABS singlets being insensitive to small magnetic perturbations. The in-gap states in the $\pi$-phase carry a local magnetic moment and react strongly on magnetic perturbations. The whole $\pi$-phase displays a Curie susceptibility diverging at zero temperature.     
\begin{figure}
\includegraphics[width=8.5cm]{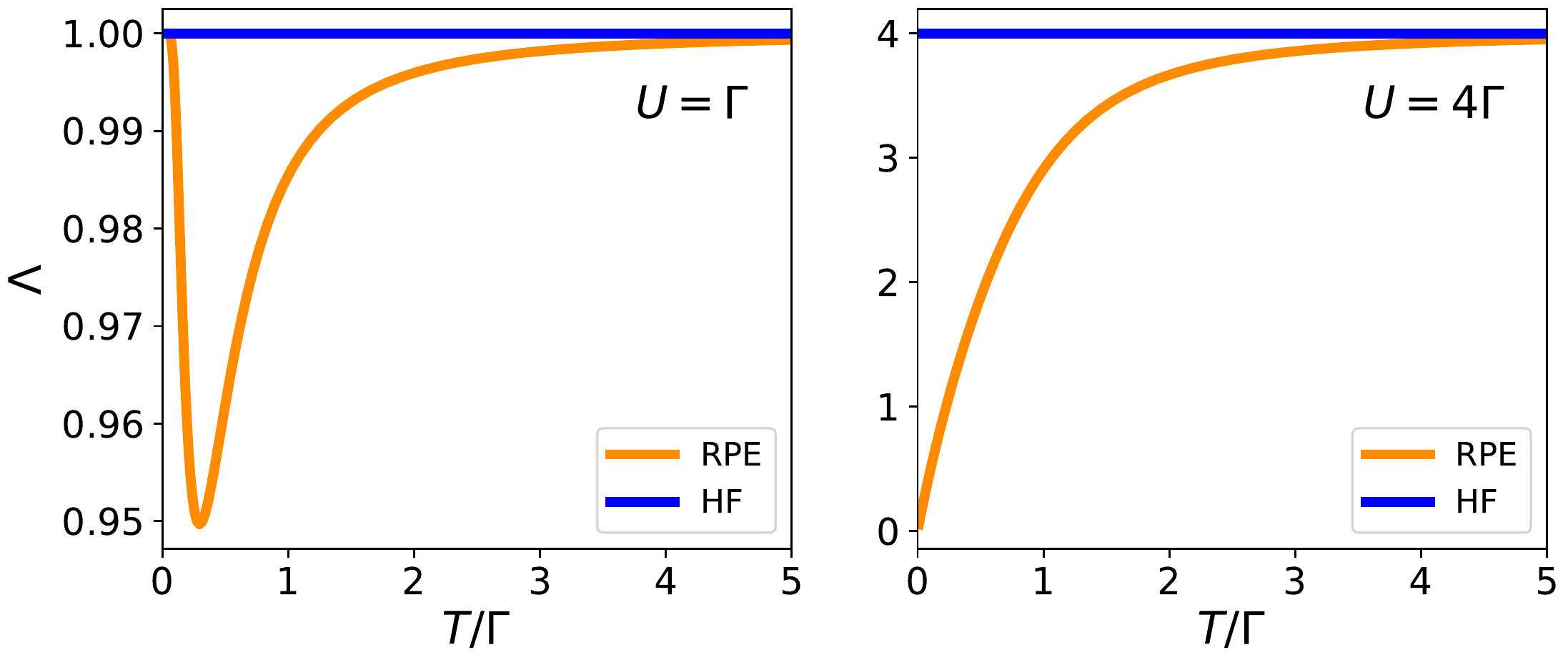}
\caption{Different temperature behavior of the vertex renormalization in the spin-symmetric state at half filling in the $0$-phase and the $\pi$-phase for the phase difference $\Phi=0$.   \label{fig:SymVertex-temp} }
\end{figure}  

The reason why the RPE suppress the HF instability is the two-particle self-consistency renormalizing the bare interaction strength $U$ to a screened effective one $\Lambda$. We plotted its temperature dependence at half filling in Fig.~\ref{fig:SymVertex-temp}. The renormalization gets stronger with the decreasing temperature but starts abating in the zero phase and dies out at zero temperature. The effective interaction approaches zero, maximizing the renormalization of the interaction, in the $\pi$-phase consistent the divergence of the magnetic susceptibility of the exact solution.      
\begin{figure}
\includegraphics[width=8.5cm]{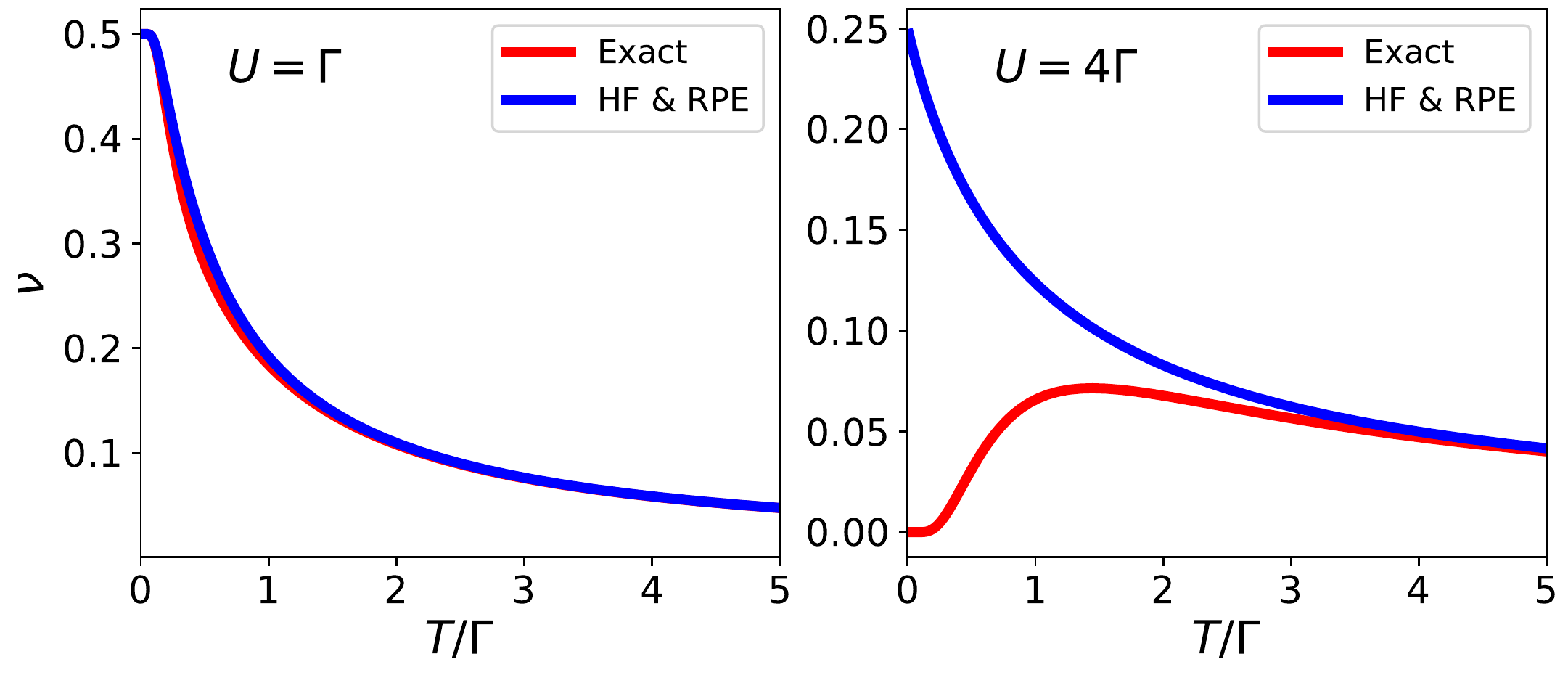}
\caption{Density of the Cooper pairs $\nu$ in the spin-symmetric state at half filling as a function of temperature  in the $0$-phase and the $\pi$-phase for the phase difference $\Phi=0$.    \label{fig:SymNu-temp} }
\end{figure}  

The thermodynamic mean-field solution with a static self-energy produces good results for quantities  with odd symmetry and sensitive to the symmetry-breaking field. It is quantitatively less accurate in determining the spin-symmetric one-particle quantities with even symmetry with respect to spin flips. We plotted the temperature dependence of the Cooper-pair density $\nu$ at half filling in the RPE/HF approximation together with the exact result in Fig.~\ref{fig:SymNu-temp}. We can see how the static spin-symmetric value deviates from the exact value at low temperatures of the $\pi$-phase. Unlike the HF mean-field, the RPE offer a direct improvement by including the dynamical corrections from the Schwinger-Dyson equation~\eqref{eq:SD-full}. It uses the renormalized interaction and this interaction is strongly renormalized at low temperatures of the $\pi$-phase. The Cooper-pair density will then better follow the dependence of the effective interaction as observed in the $0$-phase in Fig.~\ref{fig:SymVertex-temp}.  We discuss the impact of the dynamical corrections to the static self-energy in detail elsewhere.     
\begin{figure}
\includegraphics[width=7cm]{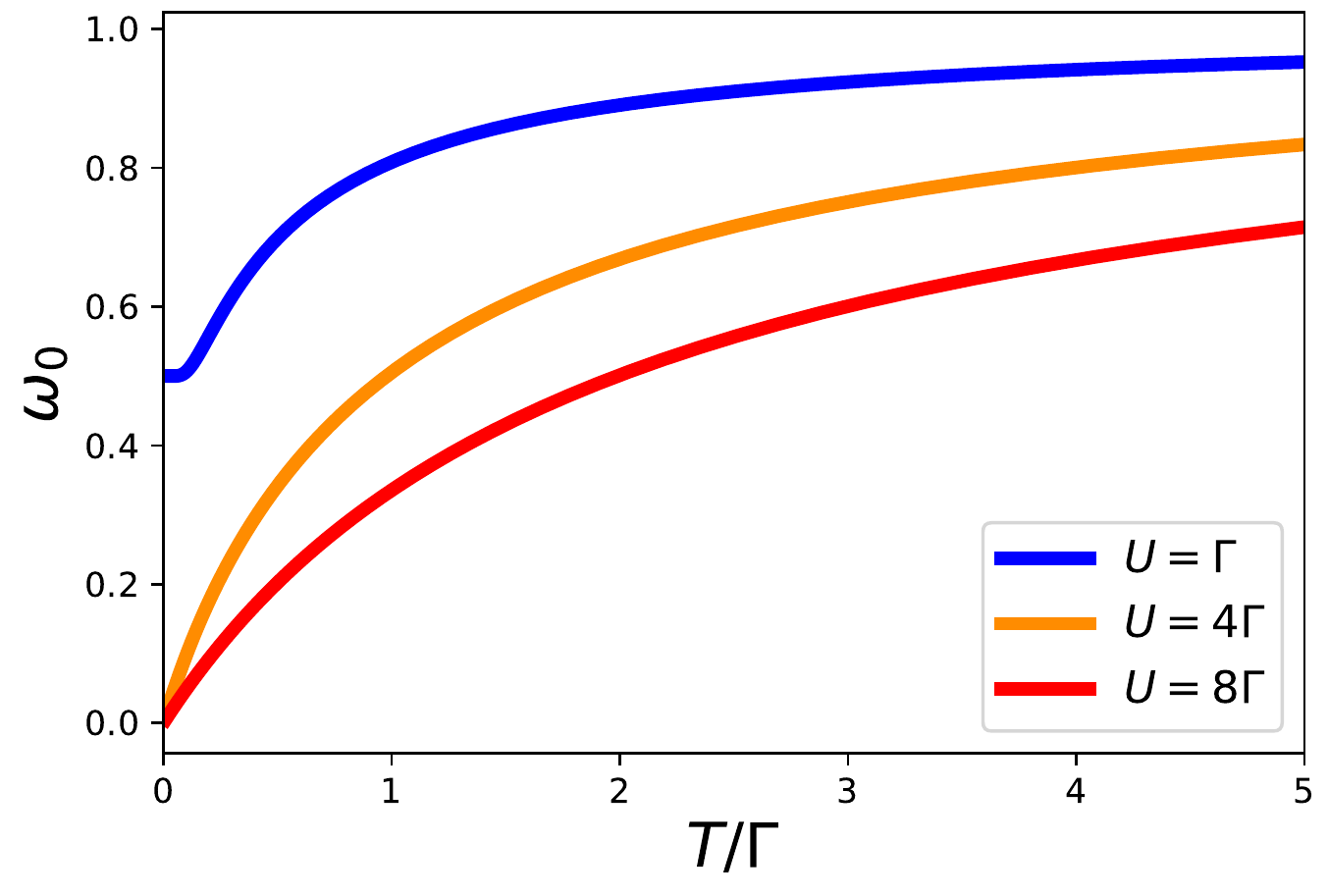}
\caption{Temperature dependence of the positive energy of the in-gap state in the spin-symmetric solution for weak and strong interactions for the phase difference $\Phi=0$.     \label{fig:InGap-Sym-temp} }
\end{figure}  

The spin-symmetric solution of the many-body Green-function approach cannot be extended to the $\pi$-phase at zero temperature since one has to cross the quantum critical point. One can nevertheless circumvent the critical point in that one extends the spin-symmetric solution to non-zero temperatures. There is no critical point at non-zero temperature and the solution can be extended continuously from weak to strong coupling \cite{Janis:2016aa}. No crossing happens and the energy of the in-gap state remains positive and approaches the quantum critical point at zero temperature, as demonstrated in Fig.~\ref{fig:InGap-Sym-temp}.  The spin-symmetric solution in the $\pi$-phase becomes unstable there and decays into the degenerate spin doublet with a saturated local magnetic moment.

\subsection{Zeeman field}

The magnetic field acting on the spin of the electrons (Zeeman field) plays an essential role in the application of the many-body Green functions in the superconducting quantum dot. It is needed to approach the zero-temperature solution in the $\pi$-phase and to see the crossing of the in-gap states at non-zero temperatures. The doublet ground state, $\pi$-phase, is degenerate and the Zeeman field is the means to lift the degeneracy. Here we analyze the properties of the low-temperature solution with an applied magnetic field.     
\begin{figure}
\includegraphics[width=8.5cm]{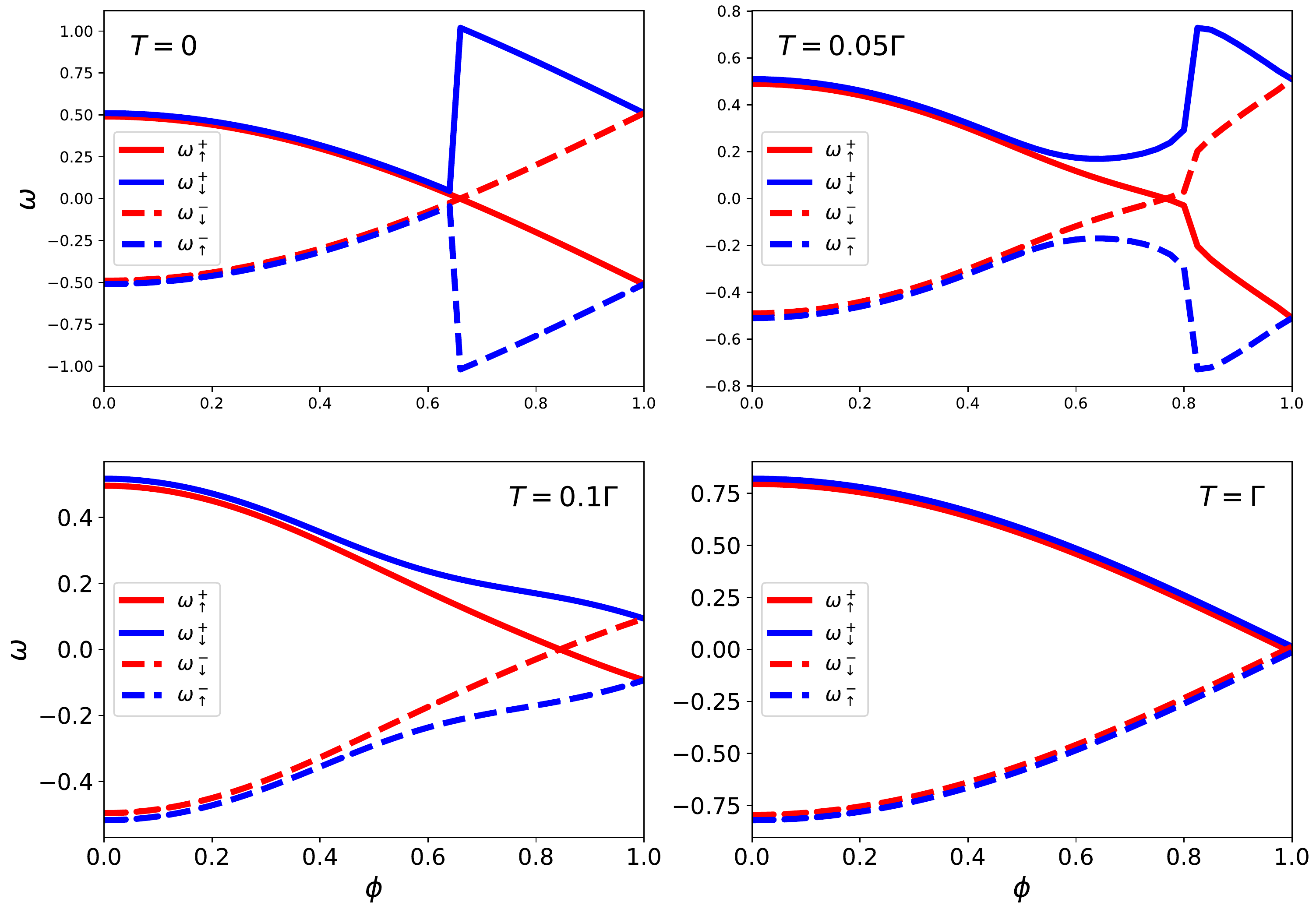}
\caption{In-gap-state energies as a function of the phase difference $\Phi$ between the superconducting leads in a weak magnetic field $h=0.01\Gamma$  for different temperatures at half filling and $U=\Gamma$. The critical angle of the crossing increases with temperature.  \label{fig:InGap-Phi} }
\end{figure}  
\begin{figure}
\includegraphics[width=8.5cm]{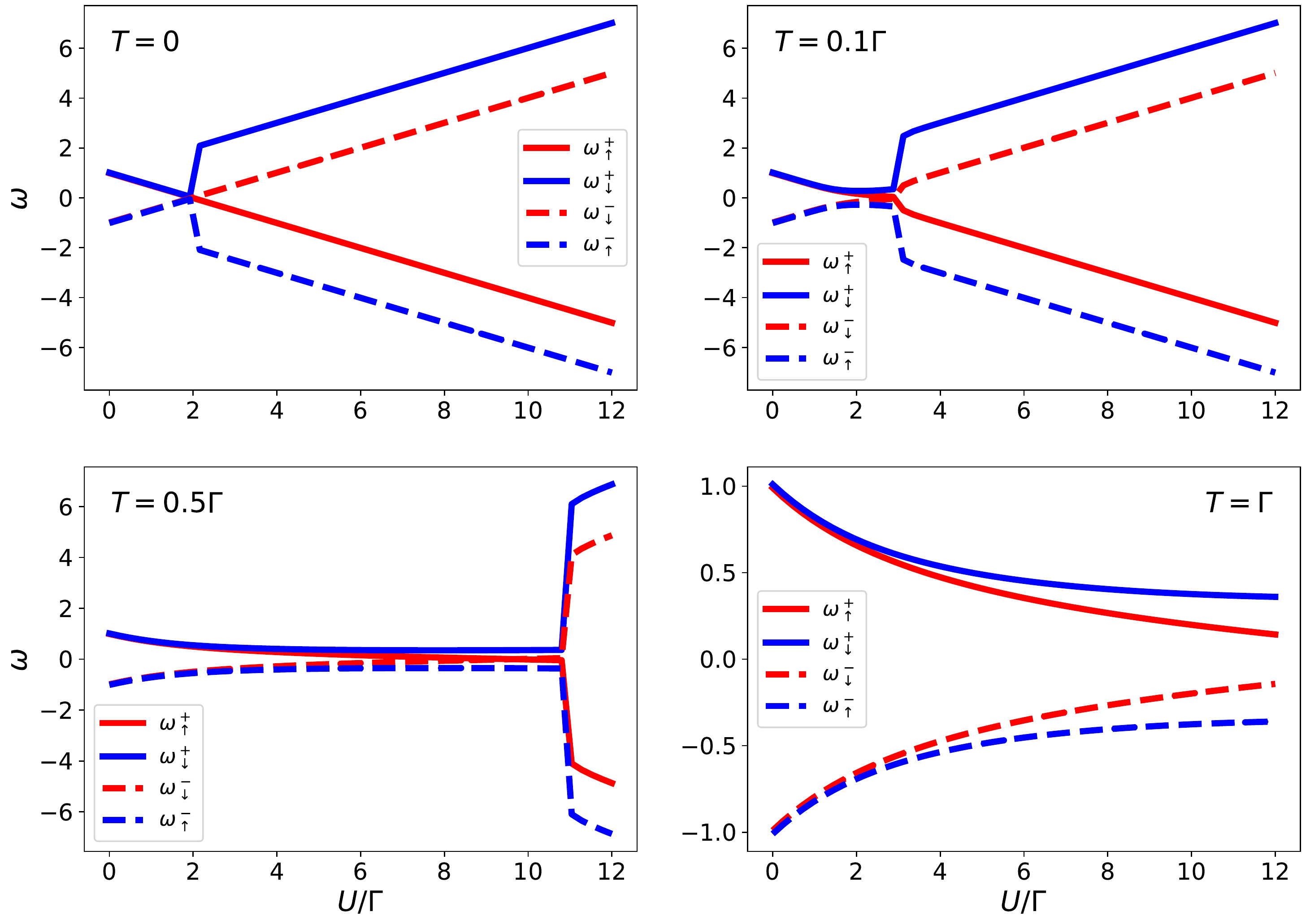}
\caption{In-gap-state energies as a function of the Coulomb repulsion $U$ in a weak magnetic field $h=0.01\Gamma$ for different temperatures at half filling and phase difference $\Phi=0$. The critical angle of the crossing increases with temperature.   \label{fig:InGap-U-temp} }
\end{figure}  

\begin{figure}
\includegraphics[width=7cm]{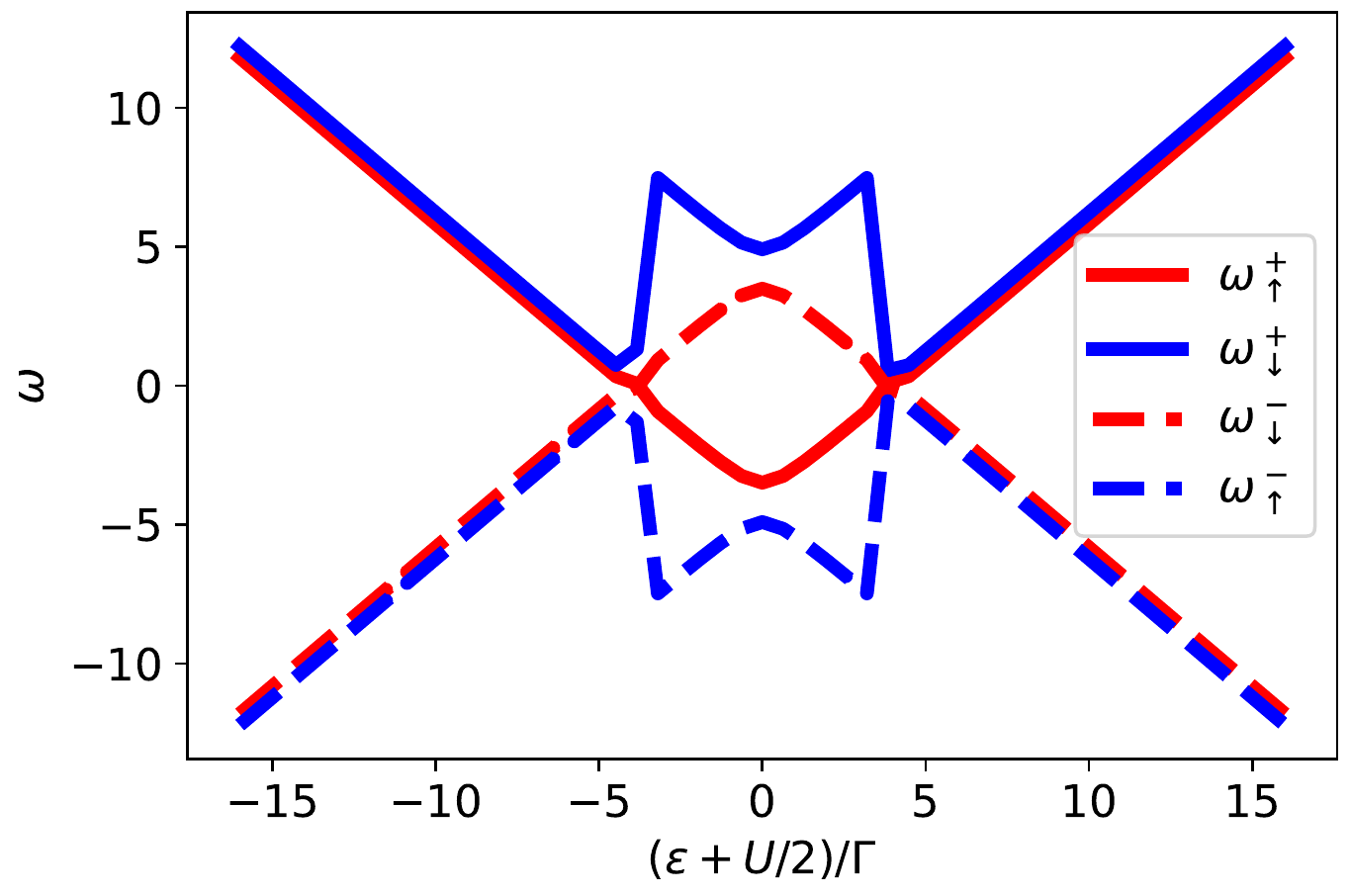}
\caption{In-gap-state energies as a function of the impurity energy level $\epsilon$ for a low temperature, $T=0.05\Gamma$, Zeeman field $h=0.2\Gamma$, interaction $U=8\Gamma$, and the phase difference $\Phi=\pi/2$ .   \label{fig:InGap-epsilon} }
\end{figure}

We plotted the dependence of the in-gap-state energies on the phase difference between the attached superconducting leads in Fig.~\ref{fig:InGap-Phi} and on the interaction strength in Fig.~\ref{fig:InGap-U-temp} for a very small magnetic field $h= 0.01\Gamma$ at different temperatures, The value of the magnetic field at the crossing increases with temperature. The curves of the in-gap-state energies are continuous due the presence of the symmetry-breaking magnetic field. We also plotted the dependence of the in-gap-state energies on the impurity energy level $\epsilon + U/2$ for strong coupling, $U=8\Gamma$ and  phase difference $\Phi=\pi/2$. We used a small magnetic field $h=0.2\Gamma$ to demonstrate the expected behavior in the $\pi$-phase, cf. Fig.~\ref{fig:InGap-epsilon}.   Note that the RPE  solution reproduces the exact positions of the in-gap states in the limit $T\to 0$ followed by the limit $h\to 0$. 

\begin{figure}
\includegraphics[width=8.5cm]{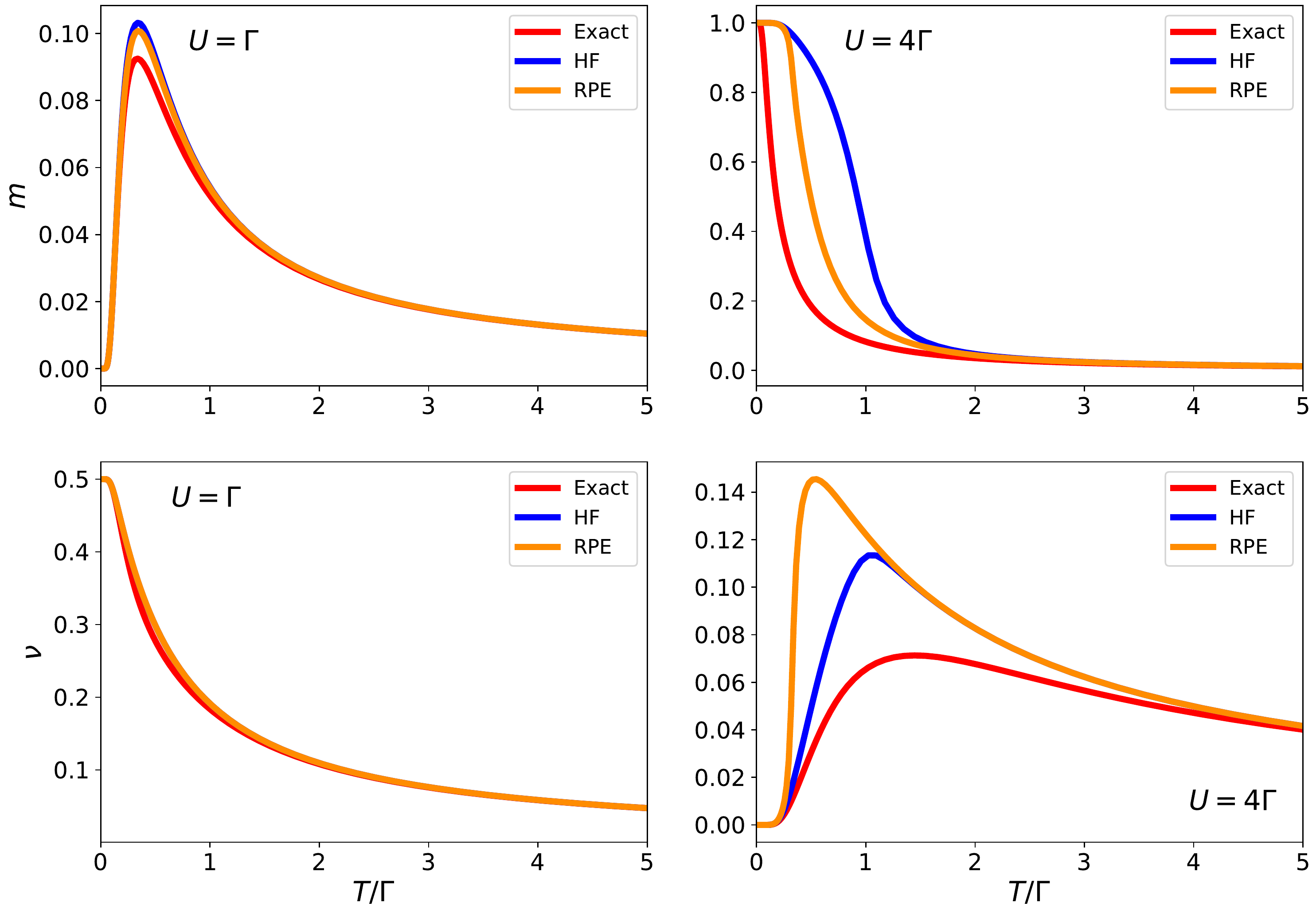}
\caption{Temperature dependence of magnetization $m$ and the Cooper-pair density $\nu$  at half filling, $h=0.1\Gamma$, and $\Phi=0$ in the $0$-phase ($U=\Gamma$)  and the  $\pi$-phase ($U=4\Gamma$) for the mean-field solutions RPE and HF compared with the exact behavior. We see that the exact behavior of the magnetization, with odd symmetry with respect to spin flip, is  better reproduced by the RPE while the Cooper-pair density, with even symmetry, is better reproduced by the HF approximation. \label{fig:MagNu-h-temp} }
\end{figure}

The low-temperature asymptotics of magnetization $m$ shows just the opposite dependence on temperature than the Cooper-pair density $\nu$ in the weak Zeeman field ($h=0.1\Gamma$) in both the $0$-phase and the $\pi$-phase, see Fig.~\ref{fig:MagNu-h-temp}. The magnetization vanishes and $\nu$ saturates in the $0$-phase in both mean-field approximations as well as in the exact one. In the $\pi$-phase the temperature asymptotics is inverted in both quantities in all solutions. The HF mean-field better simulates  the exact dependence of the Cooper-pair density $\nu$ while the RPE mean-field then better fits the exact magnetization curve.      

\begin{figure}
\includegraphics[width=8.5cm]{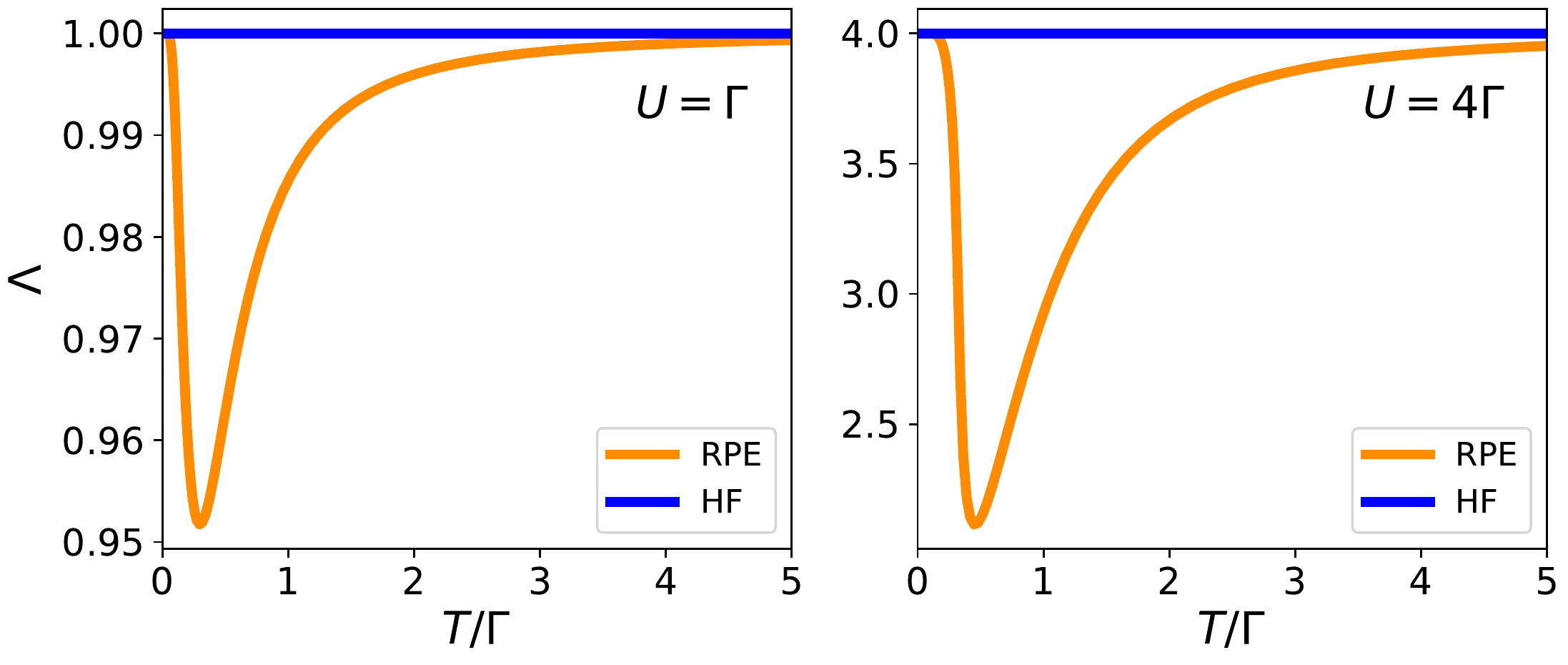}
\caption{The temperature dependence of the renormalized interaction strength $\Lambda$ at half filling in a Zeeman field $h=0.1\Gamma$, phase difference $\Phi=0$ in weak and strong couplings. The vertex $\Lambda$ shows the same rescaled dependence demonstrating that there is no difference between weak and strong interaction in the presence of the magnetic field. \label{fig:Vertex-h-temp} }
\end{figure}

The renormalized interaction $\Lambda$ in the Zeeman field has a different low-temperature asymptotics in the $\pi$-phase than in the spin-symmetric case, see Fig.~\ref{fig:Vertex-h-temp}. Once the magnetic field is kept non-zero down to zero temperature the effective interaction approaches the bare value and the exact solution is reproduced.  

\begin{figure}
\includegraphics[width=8.5cm]{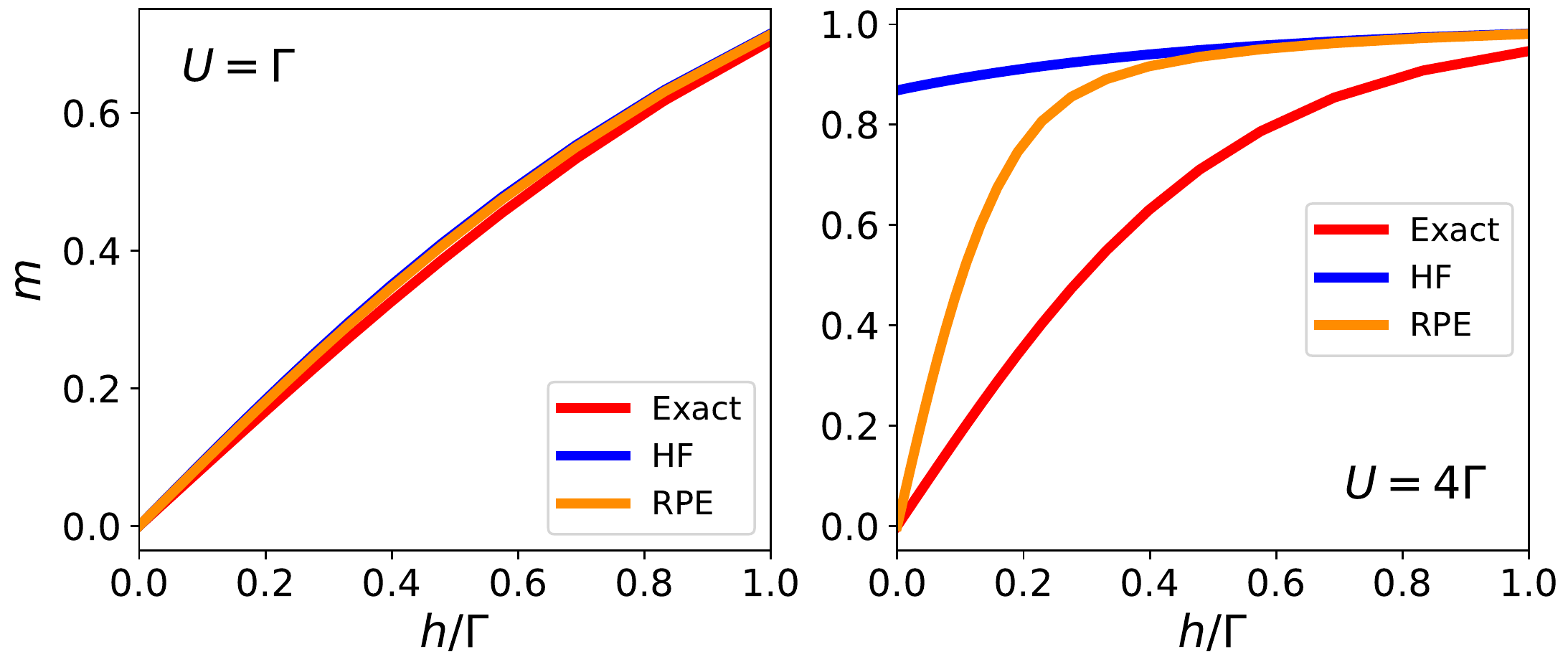}
\caption{Magnetic-field dependence of magnetization $m$ in weak ($U=\Gamma$)  and strong ($U=4\Gamma$) couplings for $\Phi=0$ and $T=0.5\Gamma$ and half filling. The strong-coupling HF solution is below its magnetic critical point  is completely off in the limit to zero magnetic field  unlike the RPE solution.      \label{fig:Mag-h} }
\end{figure}

Although the HF solution quantitatively better approximates the exact behavior of the thermodynamic quantities with even symmetry with respect to spin flips, it fails to reproduce the exact limit of the vanishing Zeeman field since it predicts a non-zero magnetization at zero field below its critical transition to the magnetic state  as documented in Fig.~\ref{fig:Mag-h}. It is our mean-field that qualitatively correctly reproduces all the limits of both one and two-particle thermodynamic quantities. 

\section{Conclusions}

The quantum dot attached to superconducting leads poses a  challenging problem to the perturbation theory with  many-body Green functions. First, electron correlations on the dot lead to a line of first-order transitions from the spin-singlet to the spin-doublet state that ends up at a quantum critical point at zero temperature where the in-gap states reach the Fermi energy.  Moreover, the doublet state is degenerate and it cannot be continuously approached from the weak-coupling spin-singlet state. The basic assumptions of the applicability of the many-body perturbation theory is a non-degenerate ground state and the existence of an analyticity region from the weak-coupling limit within which it can be applied. It cannot cope with two independent many-body equilibrium states with a first order transition between them. It may lead to a new equilibrium state, phase, only if there is a divergence accompanied by a continuous symmetry breaking. It means that the many-body perturbation expansion cannot reliably be applied to  the superconducting quantum dot at low temperatures around the quantum critical point unless a self-consistency is introduced.  A self-consistency must be introduced in the many-body perturbation expansion in order to deal with the quantum critical behavior. A static mean-field approximation  is the simplest way to achieve this goal.     

It has been known for long that the weak-coupling Hartree-Fock self-consistency is not appropriate to deal with the quantum critical point of the superconducting quantum dot since it fails at non-zero temperatures where it  leads to a spurious transition to a magnetically ordered state without the external magnetic field.  We added a two-particle self-consistency to  the HF solution  in that we replaced the bare interaction with a renormalized, screened one. We thereby suppressed the HF spurious transition to the magnetic state  and produced a fully thermodynamically consistent mean-field approximation applicable in the whole range of the input parameters. We demonstrated that it is able to deal qualitatively correctly with the quantum critical behavior of the $0-\pi$ transition in the superconducting dot as well as with the Kondo limit of the dot attached to metallic leads.

The most important finding of our mean-field theory is the manifestation of the fundamental role of the Zeeman field in the analytic description of the $0-\pi$ transition and in distinguishing the spin-singlet from the spin-doublet. The magnetic field not only lifts the degeneracy of the $\pi$-phase it allows us to determine the different character of the in-gap states in the two phases. The $0-\pi$ transition is signaled by a crossing of the energies of the in-gap states.  The in-gap states in the spin-singlet phase are the genuine Andreev bound states of two electrons with opposite spins that are insensitive to small magnetic perturbations. The low-lying excitations  in the spin-doublet phase are fermions, carry a local magnetic moment, and are sensitive  to the magnetic filed.  The magnetic susceptibility vanishes in the $0$-phase  and diverges in the $\pi$-phase at zero temperature. The equilibrium state at non-zero temperatures turns magnetic only when the Zeeman field is applied. Consequently, the weak-coupling spin-symmetric solution can continuously be extended to strong coupling at non-zero temperatures without crossing any critical point. The limit to zero field leads to a magnetic state only at zero temperature above the critical interaction strength of the $0-\pi$ transition. The limits to zero magnetic field and to zero temperature do not commute and the results depend on their order in which they are performed.  The Zeeman field  plays the role of a symmetry-breaking field we know from continuous phase transitions  and the $\pi$-phase beyond the quantum critical point mimics the ordered phase  in the lattice models.  It means that the $0-\pi$ transition happens only at zero temperature and zero magnetic field and it is a true local quantum phase transition. The crossing of the in-gap states in the magnetic field or at non-zero temperatures is noncritical with no phase transition.           

The mean-field theory presented in this paper is the first fully consistent analytic approximation that can describe not only the critical behavior of the $0-\pi$ transition but it can qualitatively correctly and reliably reproduce the behavior of the quantum dot attached to both superconducting and normal leads in the whole range of the model parameters.  It was derived within the perturbation expansion for two-particle vertices to control their critical behavior. It offers a starting point for adding dynamical corrections in a systematic way. The first step, without changing the static irreducible vertex,  is to use the Schwinger-Dyson equation to determine  the spectral properties of the model. This opens a new way to include dynamical fluctuation in the thermodynamic and spectral properties  with the controlled renormalizations of the one- and two-particle functions. 

\section*{Acknowledgment} 
Research on this problem was supported in part by Grants 19-13525S of the Czech Science Foundation. VJ thanks the INTER-COST LTC19045 Program of the Czech Ministry of Education, Youth and Sports for financial support. We thank Tom\'a\v s Novotn\'y for illuminating discussions.  

\appendix

\section{Spectral representation - Electron-hole bubble}

The thermodynamic quantities including the effective interaction can be calculated entirely in the Matsubara formalism without the necessity to continue analytically to real frequencies. If we want, however, to use the Schwinger-Dyson equation and determine the spectral properties we need a spectral representation of the two-particle bubbles, at least the electron-hole one.     
  
We decompose the imaginary part of the electron-hole bubble $\phi(\omega_{+})$ to a sum of three contributions  $\Im\phi(\omega_{+}) = \Im\phi_{bb}(\omega_{+}) + \Im\phi_{bg}(\omega_{+}) + \Im\phi_{gg}(\omega_{+})$, according to whether the arguments of the Green functions of the integrand lie both within the band, one within the band and one within the gap, and both within the gap, respectively.  We have for $\omega >0$   
\begin{widetext}
 \begin{subequations}
\begin{multline}
\Im \phi_{bb}(\omega_{+}) =    - \sum_{\sigma}\left[\int_{-\infty}^{-\Delta - \omega} + \int_{\min(\Delta - \omega, -\Delta)}^{-\Delta} + \int_{\Delta}^{\infty} \right]\frac{dx}{2\pi} \left[f(x) - f(x + \omega)\right]  \left[\Im G_{\sigma}(x_{+} + \omega) \Im G_{-\sigma}(x_{+}) 
\right. \\ \left.
 +\ \Im\mathcal{G}_{\sigma}(x_{+} + \omega) \Im \mathcal{G}_{-\sigma}(x_{+})  \right] \,,
\end{multline}
\begin{multline}
\Im \phi_{bg}(\omega_{+}) =  - \sum_{\sigma}\left[\int_{-\Delta - \omega}^{\min(\Delta - \omega, -\Delta)} + \int_{\max(\Delta - \omega, -\Delta)}^{\Delta} \right]\frac{dx}{2\pi} \left[f(x) - f(x + \omega)\right]  \left[\Im G_{\sigma}(x_{+} + \omega) \Im G_{-\sigma}(x_{+})  
\right. \\ \left.
+\ \Im\mathcal{G}_{\sigma}(x_{+} + \omega) \Im \mathcal{G}_{-\sigma}(x_{+})  \right]\,,
\end{multline}
\begin{equation}
\Im\phi_{gg}(\omega_{+}) = - \sum_{\sigma}\int_{-\Delta}^{\max(\Delta - \omega, -\Delta)}\frac{dx}{2\pi}\left[f(x) - f(x + \omega)\right] \left[\Im G_{\sigma}(x_{+} + \omega)\Im G_{-\sigma}(x_{+}) +\Im\mathcal{G}_{\sigma}(x_{+} + \omega) \Im \mathcal{G}_{-\sigma}(x_{+})  \right]\,,
\end{equation}
\end{subequations}
and for $\omega<0$
 \begin{subequations}
\begin{multline}
\Im \phi_{bb}(\omega_{+}) =    - \sum_{\sigma}\left[\int_{-\infty}^{-\Delta } + \int_{\Delta}^{\max(\Delta, -\Delta - \omega)} + \int_{\Delta - \omega}^{\infty} \right]\frac{dx}{2\pi} \left[f(x) - f(x + \omega)\right]  \left[\Im G_{\sigma}(x_{+} + \omega) \Im G_{-\sigma}(x_{+}) 
\right. \\ \left.
 +\ \Im\mathcal{G}_{\sigma}(x_{+} + \omega) \Im \mathcal{G}_{-\sigma}(x_{+})  \right] \,,
\end{multline}
\begin{multline}
\Im \phi_{bg}(\omega_{+}) =  - \sum_{\sigma}\left[\int_{-\Delta}^{\min(\Delta, -\Delta  - \omega)} + \int_{\max(\Delta, -\Delta  - \omega)}^{\Delta  - \omega} \right]\frac{dx}{2\pi} \left[f(x) - f(x + \omega)\right]  \left[\Im G_{\sigma}(x_{+} + \omega) \Im G_{-\sigma}(x_{+})  
\right. \\ \left.
+\ \Im\mathcal{G}_{\sigma}(x_{+} + \omega) \Im \mathcal{G}_{-\sigma}(x_{+})  \right]\,,
\end{multline}
\begin{equation}
\Im\phi_{gg}(\omega_{+}) = - \sum_{\sigma}\int_{\min(\Delta, -\Delta  - \omega)}^{\Delta}\frac{dx}{2\pi}\left[f(x) - f(x + \omega)\right] \left[\Im G_{\sigma}(x_{+} + \omega)\Im G_{-\sigma}(x_{+}) +\Im\mathcal{G}_{\sigma}(x_{+} + \omega) \Im \mathcal{G}_{-\sigma}(x_{+})  \right]\,.
\end{equation}
\end{subequations}
\end{widetext}
The subscript at $\omega_{+} = \omega + i0^{+}$ denotes the way the real axis is reached from the complex plane. 

The real part of the bubble is then determined from the Kramers-Kronig relation
\begin{equation}
\Re \phi(\omega) = P\int_{-\infty}^{\infty}\frac{dx}{\pi} \frac{\Im \phi(x_{+})}{x - \omega} + \phi(\infty)
\end{equation}

\section{Spectral representation - Electron-electron bubble}

The electron-electron bubble  has a simpler spectral representation. It is not needed for the spectral self-energy, but its spectral representation is useful the determination of  the effective interaction $\Lambda$ at low temperatures with a high precision. It has no  contribution from anomalous Green functions. Its imaginary part can be represented as
\begin{multline} 
\Im \psi(\omega_{+}) = - \int_{-\infty}^{\infty}\frac{dx}{\pi} \left[fx) - f(x - \omega)\right]
\\
\times\Im G_{\uparrow}(\omega_{+} - x) \Im G_{\downarrow}(x_{+})
\end{multline}
Taking into account the induced gap on the dot we can represent the three contribution from the band and gap states.
\begin{widetext}
\begin{subequations}
\begin{align}
\Im \psi_{bb}(\omega_{+}) &= - \left[\int_{-\infty}^{\min(-\Delta,\omega-\Delta)} + \int_{\max(\Delta,\omega + \Delta)}^{\infty} \right]\frac{dx}{\pi} \left[f(x) - f(x - \omega)\right] \Im G_{\uparrow}(\omega_{+} - x) \Im G_{\downarrow}(x_{+}) \,,
\\
\Im \psi_{bg}(\omega_{+}) &= - \left[\int_{\min(-\Delta,\omega-\Delta)}^{\max(-\Delta,\omega-\Delta)} + \int_{\min(\Delta,\omega+\Delta)}^{\max(\Delta,\omega+\Delta)} \right]\frac{dx}{\pi} \left[f(x) - f(x - \omega)\right] \Im G_{\uparrow}(\omega_{+} - x) \Im G_{\downarrow}(x_{+}) \,,
\\
\Im \psi_{gg}(\omega_{+}) &= - \int_{\max(-\Delta,\omega-\Delta)}^{\min(\Delta,\omega+\Delta)} \frac{dx}{\pi} \left[f(x) - f(x - \omega)\right] \Im G_{\uparrow}(\omega_{+} - x) \Im G_{\downarrow}(x_{+}) \,.
\end{align}
\end{subequations}
\end{widetext}

\section{Atomic limit - Exact solution}

We summarize the basic results of the exact solution of the atomic limit with infinite superconducting gap. The atomic Hamiltonian is a matrix  
\begin{equation}\label{eq: effective Hamiltonian of SQDS-INF -- matrix form}
H = 
\left(
\begin{matrix}
0 & 0 & 0 & \Gamma c_\Phi\\
0 & \epsilon_d+h & 0 & 0\\
0 & 0 & \epsilon_d-h & 0\\
\Gamma c_\Phi & 0 & 0 & 2\epsilon_d+U
\end{matrix} 
\right) \,.
\end{equation}

One can simply diagonalize the Hamiltonian matrix, Eq.\eqref{eq: effective Hamiltonian of SQDS-INF -- matrix form}, by observing that the central $2\times2$ sub-block is decoupled from the others.
As a result, the $4$ eigenstates are summarized in the following:
(i) $E_d^- = \epsilon_d - h$ with the eigenstates $\ket{1,0}$;
(ii) $E_d^+ = \epsilon_d + h$ with the eigenstates $\ket{0,1}$;
(iii) $E_s^- = \frac{1}{2}\left[ 2\epsilon_d + U - \sqrt{(2\epsilon_d+U)^2 + 4\Gamma^2 c_\Phi^2} \right]$ with the eigenstates $\frac{1}{\sqrt{\Gamma^2 c_\Phi^2 + (E_s^-)^2}} \left( \Gamma c_\Phi \ket{0,0} + E_s^-\ket{1,1} \right)$;
and (iv) $E_s^+ = \frac{1}{2}\left[ 2\epsilon_d + U + \sqrt{(2\epsilon_d+U)^2 + 4\Gamma^2 c_\Phi^2} \right]$ with the eigenstates $ \frac{1}{\sqrt{\Gamma^2 c_\Phi^2 + (E_s^+)^2}} \left( \Gamma c_\Phi \ket{0,0} + E_s^+\ket{1,1} \right) $.

The general thermodynamic quantity is 
\begin{equation}\label{eq: physical quantity formal definition}
Q = \frac{1}{Z} \sum_i \avg{E_i |\hat{Q}| E_i} e^{-\beta (E_i - \mu N_i)}\,,
\end{equation}
where  $E_i$ and $\ket{E_i}$ are the eigenvalues and corresponding eigenstates of the Hamiltonian,  $\beta = 1/k_{B}T$, and the partition function is
\begin{equation}\label{eq: partition function}
\begin{split}
Z = e^{-\beta E_s^-} + e^{-\beta E_s^+} + e^{-\beta (\epsilon_d + h)} + e^{-\beta(\epsilon_d - h)} \,.
\end{split}
\end{equation}

The thermodynamic properties can, alternatively, be derived from the derivatives of the grand potential $F = -\frac{1}{\beta} \ln Z$. The charge and spin densities are 
\begin{subequations}
\begin{multline}
n = \frac{1}{Z} \left[ \sum_\sigma e^{-\beta(\epsilon_d - \sigma h)} + \frac{2(E_s^-)^2}{\Gamma^2c^2_\phi + (E_s^-)^2}e^{-\beta E_s^-} 
\right. \\ \left.
+\ \frac{2(E_s^+)^2}{\Gamma^2c^2_\phi + (E_s^+)^2}e^{-\beta E_s^+} \right] \,,
\end{multline}
\begin{equation}
\begin{split}
	m = \frac{1}{Z}\left[ \sum_\sigma \sigma e^{-\beta(\epsilon_d - \sigma h)}\right]
\end{split} \,.
\end{equation}
\end{subequations}
The density of the Cooper pairs is 
\begin{multline}
\nu =  \frac{1}{Z}\left[ \frac{\Gamma  E_s^-}{\Gamma^2 c^2_\Phi + (E_s^-)^2} e^{-\beta E_s^-} 
\right. \\ \left. 
+\ \frac{\Gamma  E_s^+}{\Gamma^2 c^2_\Phi + (E_s^+)^2} e^{-\beta E_s^+} \right] \,.
\end{multline}

\bibliographystyle{apsrev4-1}
%

\end{document}